\documentclass[letterpaper,twocolumn,10pt]{article}
\usepackage{usenix2019_v3}

\usepackage{amsmath}
\usepackage{breakurl}
\usepackage{cancel}
\usepackage{color}
\usepackage{enumitem}
\usepackage{graphicx}
\usepackage{listings}
\usepackage{nicefrac}
\usepackage{siunitx}
\usepackage{subcaption}
\usepackage{url}

\graphicspath{{badges/}}
\usepackage[available,functional,reproduced]{usenixbadges}

\usepackage{titlesec}
\titlespacing*\section{0pt}{4pt plus 2pt minus 2pt}{2pt plus 1pt minus 1pt}
\titlespacing*\subsection{0pt}{3pt plus 2pt minus 1pt}{1pt plus 1pt minus 1pt}
\titlespacing*\subsubsection{0pt}{2pt plus 2pt minus 1pt}{1pt plus 1pt minus 1pt}

\usepackage{titling}
\usepackage{caption}
\DeclareMathAlphabet{\mathcal}{OMS}{cmsy}{m}{n}

\newif\iffull

\newcommand{\secref}[1]{\S{}\ref{#1}}
\newcommand{\tabref}[1]{Table \ref{#1}}
\newcommand{\figref}[1]{Fig. \ref{#1}}

\newcommand{\parhead}[1]{\noindent\textbf{#1.}}

\newcommand{\sys}{MAGE}
\newcommand{\bigO}{\mathcal{O}}
\newcommand{\gc}{$\widetilde{C}$}
\newcommand{\gates}{G}
\newcommand{\wires}{W}
\newcommand{\inputs}{I}
\newcommand{\outputs}{O}
\newcommand{\swapin}{\textsc{Swap-In}}
\newcommand{\swapout}{\textsc{Swap-Out}}
\newcommand{\iswapin}{\textsc{Issue-Swap-In}}
\newcommand{\iswapout}{\textsc{Issue-Swap-Out}}
\newcommand{\fswapin}{\textsc{Finish-Swap-In}}
\newcommand{\fswapout}{\textsc{Finish-Swap-Out}}
\newcommand{\NP}{\text{NP}}

\newcommand{\merge}{\textbf{merge}}
\newcommand{\sort}{\textbf{sort}}
\newcommand{\ljoin}{\textbf{ljoin}}
\newcommand{\mvmul}{\textbf{mvmul}}
\newcommand{\binfclayer}{\textbf{binfclayer}}

\newcommand{\rsum}{\textbf{rsum}}
\newcommand{\rstats}{\textbf{rstats}}
\newcommand{\rmvmul}{\textbf{rmvmul}}
\newcommand{\naivermatmul}{\textbf{n\_rmatmul}}
\newcommand{\tiledrmatmul}{\textbf{t\_rmatmul}}

\begin{document}

\date{}

\title{\Large \bf \sys{}: Nearly Zero-Cost Virtual Memory for Secure Computation}

\author{
{\rm Sam Kumar, David E. Culler, and Raluca Ada Popa}\\
\textit{University of California, Berkeley}
} 

\maketitle


\begin{abstract}
Secure Computation (SC) is a family of cryptographic primitives for computing on encrypted data in single-party and multi-party settings.
SC is being increasingly adopted by industry for a variety of applications.
A significant obstacle to using SC for practical applications is the memory overhead of the underlying cryptography.
We develop \sys{}, an execution engine for SC that efficiently runs SC computations that do not fit in memory.
We observe that, due to their intended security guarantees, SC schemes are inherently \emph{oblivious}---their memory access patterns are independent of the input data.
Using this property, \sys{} calculates the memory access pattern ahead of time and uses it to produce a memory management plan.
This formulation of memory management, which we call \emph{memory programming}, is a generalization of paging that allows \sys{} to provide a highly efficient virtual memory abstraction for SC.
\sys{} outperforms the OS virtual memory system by up to an order of magnitude, and in many cases, runs SC computations that do not fit in memory at nearly the same speed as if the underlying machines had \emph{unbounded physical memory} to fit the entire computation.

\end{abstract}

\section{Introduction}\label{sec:introduction}

Secure Computation (SC) refers to cryptographic primitives that allow computation on encrypted data.
An example of SC is secure multi-party computation, which allows two parties to perform a collaborative computation on private input data.
Advances in cryptography over the years have steadily brought SC closer to practice.
Recently, the use of SC in industry---in particular, homomorphic encryption (HE) and secure multi-party computation (SMPC)---has burgeoned.
Companies offer services based on SC~\cite{inpher, duality, unbound, cape, curv, keyless} (from secure collaborative learning to decentralized authentication and custody), large financial enterprises have added SC-based products~\cite{ant}, and cryptocurrencies secure billions of dollars with SC~\cite{zcash}.

SC not only has high CPU overhead, but also requires high memory usage and, in the case of SMPC, high network usage.
For example, a 64-bit integer, which requires only 8~B of memory when computing in plaintext, takes up  1~KiB of memory when using a garbled circuit (a type of SMPC).
Efficiently running SC requires careful attention to not only CPU efficiency, but also memory and network demands.

CPU overheads can be reduced using hardware accelerators (e.g., GPUs, FPGAs) or specialized hardware (e.g., AES-NI).
Network bandwidth continues to grow exponentially according to Nielsen's Law~\cite{nielsen1998law}, and recent cryptographic improvements have relaxed network bandwidth demands for some SC protocols~\cite{boyle2019efficient, chen2020maliciously}.
But memory management remains problematic.
Many recent cryptographic systems based on SC report that SC systems quickly run out of memory~\cite{volgushev2019conclave, poddar2021senate, zheng2019helen, zhu2018nanopi}.
Once they do, the computation becomes prohibitively slow because the OS inefficiently swaps the large memory footprint to secondary storage.
For example, the authors of Conclave~\cite{volgushev2019conclave} report that Obliv-C, an SMPC framework, can run a join on only 30,000 records before running out of memory, and state that SMPC ``in practice only scales to a few thousand input records.''
Similarly, Senate~\cite{poddar2021senate}, a secure collaborative analytics engine based on SMPC, can run a 16-party private set intersection on only 10,000 integers per party.

In this context, we address the research question: \textbf{Can SC execute efficiently when it does not fit in memory?}
We answer this in the affirmative with our system \sys{}.\footnote{\sys{} stands for \textbf{M}emory-\textbf{A}ware \textbf{G}arbling \textbf{E}ngine.}

A natural starting point for \sys{} is to specialize the OS page replacement policy to SC workloads.
Unsurprisingly, this design suffers from some of the same pitfalls as classic virtual memory systems.
Pages may not be fetched until a page fault occurs, requiring multiprogramming to keep the CPU busy~\cite{denning1980workingsets}.
Furthermore, classic page replacement algorithms perform poorly on some workloads~\cite{belady1966study}, and a policy specialized to SC would likely be no different.

To mitigate these issues, we observe that SC is inherently \emph{oblivious}.
In particular, many SC protocols have \emph{no data-dependent memory accesses}.
This is because an SC protocol must not leak any information about the data contents via its memory access pattern.
Our key insight in \sys{} is that SC's inherent obliviousness allows us to calculate the access pattern for any computation \emph{in advance} and use it to manage memory in a fundamentally more efficient way than classic OS paging.
Unlike paging, which typically responds to page faults \emph{reactively}, \sys{} can \emph{proactively} produce a memory management plan based on the program's memory access pattern.
To highlight this distinction, we call our approach \emph{memory programming} and the resulting plan a \emph{memory program}.
\sys{} preplans the exact sequence of memory-storage transfers to issue at runtime, given a target memory consumption.
Enabled by memory programming and the compute-to-memory ratio of SC workloads, \sys{} executes certain SC programs that do not fit in memory at nearly in-memory speeds, as if memory were unbounded---in effect, virtual memory at nearly zero cost.

To understand the power of \sys{}'s preplanning based on SC's obliviousness, consider Belady's theoretically optimal paging algorithm (MIN)~\cite{belady1966study}.
MIN, being a clairvoyant algorithm, is not realizable in the classic formulation of paging; it is typically used as a point of comparison to other realizable heuristics.
But in the context of memory programming, \sys{} can use MIN directly, because it knows the access pattern in advance.
Memory programming allows \sys{} to use an algorithm that is well-grounded in theory, instead of a heuristic (e.g., LRU or LFU) that sometimes performs poorly.

Yet memory programming also raises the bar for possible memory management strategies.
For example, although MIN is an optimal paging algorithm, it unfortunately does not produce an optimal memory program.
The reason is that MIN, like other paging algorithms, brings a page into memory only at the moment it is needed, thereby causing the program to stall while transferring the page.
We can overcome this by leveraging SC's obliviousness once again, to prefetch according to the access pattern (i.e., with no false positives or false negatives) so that the program never stalls.

At its core, our approach to memory management is quite simple: \sys{} optimizes storage bandwidth by evicting pages using MIN, and optimizes latency via prefetching and asynchronous eviction.
Whereas classic paging algorithms typically rely on heuristics and empirical observations of what works well in practice~\cite{bovet2006pfra}, our memory programming approach is simple, well-grounded, robust, and performant.

While conceptually simple, the above strategy is challenging to instantiate efficiently.
The reason is that MIN requires the entire memory access pattern to be materialized at once; it cannot be applied in a streaming fashion.
Using Intel Pin~\cite{luk2005pin}, we found that an SC workload that runs in under an hour can issue \emph{trillions} of memory accesses.
Thus, materializing the access trace could require \emph{terabytes} of space.

To address this, we leverage the  strong precedent for using DSLs to specify SC programs~\cite{hastings2019compilers, viand2021compilers}.
\sys{}'s planner represents the program as a bytecode recording higher-level operations specified in the DSL program.
This is more succinct than recording individual memory accesses.
For example, consider a program that adds two integers using garbled circuits, an SMPC protocol.
Garbled circuits support only AND and XOR operations on encrypted \emph{bits}, so the integer addition is ultimately decomposed into encrypted AND and XOR operations, each of which comprises many memory accesses.
Yet, \sys{} records the entire addition operation as a \emph{single} entry in the bytecode.
This works well because most of the addition operation's memory accesses are ``uninteresting''---they are accesses to temporary variables (e.g., on the stack) that fit easily in memory, or to SC protocol state that should remain in memory for the entire program.
The only consequential accesses for memory management---reading the two input integers and writing the output integer---are captured in the single entry \sys{} records.

Once \sys{} allows SC to efficiently expand beyond the physical memory limit, another limited resource (e.g., storage/network bandwidth or CPUs) of a single machine could become the bottleneck.
Thus, we design \sys{} to support \emph{parallel} SC execution across multiple network flows, CPU cores, or machines.
To do so, we observe that a distributed memory programming model allows SC to be parallelized in this way, without requiring \sys{}'s planner to reason about threads executing concurrently in the same address space.

Finally, we aim to support a variety of applications and protocols, including new ones that may emerge in the coming years.
The challenge is that different SC protocols may be very different cryptographically and may support different operations efficiently.
Fortunately, our memory programming approach allows us to build \sys{} entirely in userspace on a Linux system, helping to make \sys{} \emph{extensible} to new applications and protocols.
We carefully design a layered architecture for \sys{} so that the DSL, bytecode, and interpreter can be extended for new SC protocols.

We implemented \sys{} in C++ and apply it to two SC protocols: (1) garbled circuits, a type of SMPC, and (2) CKKS, a type of HE.
We evaluated \sys{} using 10 workloads, sized such that they do not fit in memory.
\sys{} outperforms the operating system's virtual memory for all 10 workloads, and outperforms it by $4$--$12\times$ for 7 of them.
Additionally, \sys{} executes all 10 workloads at within $60\%$ of in-memory speeds, and runs 7 of them at within 15\% of in-memory speeds.

Even with our techniques, SC remains orders of magnitude slower than plaintext computation due to CPU and network overheads.
That said, various applications like federated data analytics~\cite{bater2017smcql, poddar2021senate}, coopetitive machine learning~\cite{zheng2019helen}, and privacy-preserving recommendation~\cite{nikolaenko2013privacy} \emph{require} SC.
Due to privacy constraints, running these applications in plaintext is not an option.
By bringing memory management overhead for SC to nearly zero, \sys{} helps make SC more practical and potentially enables more SC-based applications.

\section{Secure Computation Background}\label{sec:background}

\subsection{Circuit Representation}
As explained earlier, SC is inherently oblivious, meaning that any function $f$ computed using SC cannot have data-dependent memory accesses.
Thus, it is natural to describe the function $f$ as a circuit $C$~\cite{malkhi2004fairplay, huang2011faster, carpov2015armadillo, dathathri2020eva}.
$C$ is a combinational circuit that accepts the arguments to $f$ as inputs and produces the result of $f$ applied to those arguments as its output.
We write $C = (\wires, \gates)$, where $\wires$ is a set of wires and $\gates$ is a set of gates.
Each \emph{wire} represents a datum whose type is the unit of computation in the SC scheme (in most cases, it is the information stored in a single ciphertext).
We denote the subset of $\wires$ storing $C$'s input as $\inputs$, and the subset of $\wires$ storing $C$'s output as $\outputs$.
Each \emph{gate} represents a computation supported by the SC scheme.
We will typically assume that each gate has exactly one output wire, and that each $w\not\in\outputs$ is the input wire of at least one gate.
Thus, $|\wires| = |\gates| + |\inputs|$.

The particular data types represented in the wires and the types of supported gates depend on the particular SC scheme of interest.
For the CKKS homomorphic encryption scheme~\cite{cheon2017homomorphic}, each wire represents a \emph{vector of real numbers} and each gate represents an element-wise \emph{addition or multiplication} of those vectors.
For garbled circuits~\cite{yao1986secrets}, each wire represents a \emph{single bit} and each gate represents a binary \emph{AND operation or XOR operation} on those bits.
Other SC schemes can be similarly formulated this way.
Below, we explain CKKS and garbled circuits in greater depth.

\subsection{CKKS Homomorphic Encryption}
In the CKKS scheme~\cite{cheon2017homomorphic}, each ciphertext encodes a vector of real or complex numbers (stored with limited precision).
Given ciphertexts $c_1 = \textsf{Enc}(\vec{v_1})$ and $c_2 = \textsf{Enc}(\vec{v_2})$, one can compute $\textsf{Enc}(\vec{v_1} + \vec{v_2})$ and $\textsf{Enc}(\vec{v_1} \circ \vec{v_2})$ (where $\circ$ is element-wise multiplication).
The dimension of each vector depends on parameters chosen during key generation.
Each ciphertext is assigned a level, which is a nonnegative integer.
When performing the element-wise multiplication operation, both input ciphertexts must have the same level; the level of the output ciphertext is one less than the level of the inputs.
Performing element-wise addition does not reduce the ciphertext level the way element-wise multiplication does.
A ciphertext at level 0 cannot be used for element-wise multiplication.
The maximum level of a ciphertext depends on the parameters chosen during key generation.
While one can run a bootstrapping procedure to increase the level of a ciphertext, it is very expensive, and therefore not implemented by all libraries.

\subsection{Garbled Circuits}\label{s:garbled_circuits}
Yao's garbled circuit protocol~\cite{yao1986secrets} (referred to simply as \emph{garbled circuits}) allows two parties, called the \emph{garbler} and the \emph{evaluator}, to jointly compute a function $f$ over their private inputs $x_1$ and $x_2$.
The protocol requires $f$ to be represented as a \emph{boolean circuit} $C$.
Unlike CKKS, there are no restrictions on $C$'s depth.
However, \emph{both} parties have to execute the circuit.

First, the two parties run a protocol called \emph{oblivious transfer} to obtain the (encrypted) wire values for their inputs without revealing their inputs.
Then the garbler encrypts $C$ in a special way called \emph{garbling} to obtain \gc{}, called a \emph{garbled circuit}.
The process of garbling is analogous to executing the circuit; a gate cannot be garbled until the (encrypted) values of both input wires are obtained, and garbling a gate produces, as a side effect, the (encrypted) value of the output wire.
Then, the garbler sends \gc{} to the evaluator.
The evaluator executes the circuit, executing each gate using the gate's garbled information in \gc{}.
Finally, the two parties communicate to decipher the plaintext values of the output wires.

If the parties would like to repeat the computation again with different inputs, they must re-garble $C$.
It is insecure to reuse the same garbled circuit \gc{} with different sets of inputs.

More comprehensive explanations of garbled circuits, their underlying cryptography, and their state-of-the-art optimizations are available in other resources~\cite{rosulek2015history, bicer2017yao, yakoubov2017gentle}.

\subsection{Efficiently Executing Circuits}\label{s:memmgmt}
In this section, we give background on existing techniques for efficiently executing cryptographic circuits.
Although many of these techniques were developed for garbled circuits, they mostly apply to homomorphic encryption as well.

\subsubsection{Na\"ive Baseline}

Early garbled circuit systems, like Fairplay~\cite{malkhi2004fairplay}, JKS~\cite{jha2008genomic}, and PSPW~\cite{pinkas2009practical}, allocate memory for all wires and store the entire garbled circuit in memory.
The memory overhead is $\bigO(|\wires| + |\gates|)$.
Because, for a well-formed circuit, $|\gates| + |\inputs| = |\wires|$, this is equivalent to $\bigO(|\wires|)$.

\subsubsection{Pipelining Garbling and Evaluation}\label{s:streaming}
After the garbler garbles a gate to include in \gc{}, the garbler does not use that gate's garbled data.
Similarly, once the evaluator evaluates a gate, it never again uses that garbled gate.
Based on this observation, the HEKM system~\cite{huang2011faster} operates without keeping the entire garbled circuit in memory, as follows.
The garbler and the evaluator first agree on an order in which to execute the gates in $C$.
Then, the garbler garbles each gate and streams the garbled gates to the evaluator, who evaluates the gates in the same order.
In this way, all gates are garbled and evaluated, without materializing the full set of garbled gates at any one time.
Because space is allocated for all wires in the circuit, the memory overhead is still $\bigO(|\wires|)$.

\subsubsection{Reclaiming Wire Memory}\label{s:wire_reuse}
When executing a circuit, one can discard the memory for a wire once all gates it feeds into have been executed.
Only wires whose values have been computed and will be used in the future---the \emph{live} wires---must be kept in memory.
The KSS system~\cite{kreuter2012billion} takes advantage of this by dynamically attaching a reference count to each wire; PCF~\cite{kreuter2013pcf} statically calculates when to reuse wire memory.
Using interpretation techniques developed in PCF~\cite{kreuter2013pcf} and refined in Frigate~\cite{mood2016frigate}, not even the plaintext circuit is materialized in memory.
TinyGarble~\cite{songhori2015tinygarble}, EMP-toolkit~\cite{emp-toolkit} (for semi-honest SMPC), and EVA~\cite{dathathri2020eva} also use variants of this technique.
With this optimization, the memory demand is $\bigO(w)$, where $w$ is the size of the largest set of live wires when executing the circuit.
\sys{} builds on this line of work by exploring how to efficiently swap to storage when $w$ wires do not fit in memory.

\section{Memory Overhead of Secure Computation}\label{sec:scalability}
First, we discuss the memory overhead of SC.
Then, we discuss the memory overhead for collaborative applications.

\subsection{Analysis of the Memory Demand}\label{s:mem_analysis}
The size of the circuit, for a computation, is proportional to the size of the computation.
But in many cases, the memory demand is substantially smaller than the circuit size; only $w$ wires need to be stored, where $w$ is the size of the largest set of live wires when executing the circuit (\secref{s:wire_reuse}).

In practice, circuits are often described in a programming language ~\cite{hastings2019compilers, viand2021compilers} and gates are executed in the same order as the program is interpreted.
In this execution order, live wires correspond to in-scope variables in the program.
Thus, the memory usage of running an SC program has the same order of growth as running the same algorithm in plaintext.

The memory cost of SC lies in the constant factors.
When executing a secure computation protocol, the wire values are encrypted. Thus, a key parameter is the expansion factor of the encryption.
In garbled circuits using a 128-bit block cipher, including state-of-the-art optimizations (Point-and-Permute~\cite{beaver1990round}, Free XOR~\cite{kolesnikov2008freexor},
Half Gates~\cite{zahur2015halfgates}, and Fixed-Key Block Cipher~\cite{bellare2013fixedkey, guo2019mitccrh}), each wire value is 16 bytes.
Each wire represents only 1 bit of plaintext, so this is a $128\times$ expansion factor.
For CKKS, ciphertexts at higher levels are larger than ciphertexts at lower levels.
For the parameters we used in our evaluation, each ciphertext is hundreds of kilobytes and encodes a vector of dimension up to 4,096.

\subsection{Scaling Collaborative Applications}\label{s:minimal}
SMPC supports \emph{collaborative applications} over secret data, such as federated data analytics~\cite{bater2017smcql} and cooperative machine learning~\cite{mohassel2017secureml}.
A common technique to reduce SMPC's overhead is to use SMPC in a \emph{minimal way}.
For example, some approaches aim to use SMPC for only a small part of the overall computation~\cite{bater2017smcql, liu2017minionn, juvekar2018gazelle, volgushev2019conclave, zheng2019helen}.
Others carefully choose algorithms that can be executed efficiently in SMPC or use approximations that incur less overhead~\cite{mohassel2017secureml, riazi2019xonn, mishra2020delphi}.
But even with these approaches, the SMPC computation often has high memory demands~\cite{poddar2021senate}.
Thus, it remains important to efficiently execute SMPC computations that do not fit in memory.

\section{Overview of \sys{}}\label{sec:overview}

\begin{figure}[t]
    \centering
    \includegraphics[width=\linewidth]{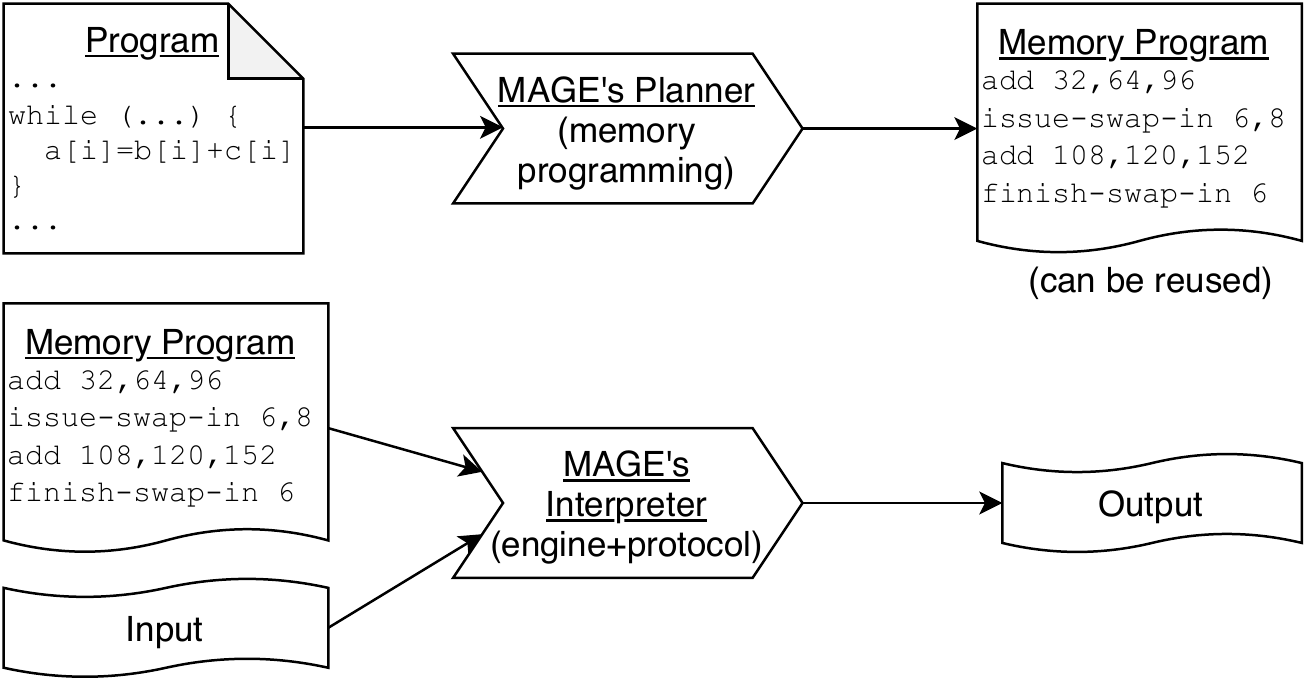}
    \caption{Overview of \sys{}. It consists of two phases: planning (top) and execution (bottom)}
    \label{fig:overview}
\end{figure}

SC workloads are oblivious by nature.
Thus, \sys{} can work out the program's memory access pattern in advance, and use this information to produce a memory management plan, called a \emph{memory program}, tailored to the particular access pattern.
Importantly, obliviousness is not merely an artifact of certain existing SC schemes; it is inherent to SC.
Otherwise, an adversary could potentially infer information about secret data based on the memory access pattern.

To support this paradigm, \sys{}'s workflow has two phases, as shown in \figref{fig:overview}.
An SC application is written in a DSL internal to C++.
\sys{}'s planner unrolls the DSL code to produce a bytecode, and then performs transformations on the bytecode to produce a memory program.
In \sys{}, the memory program is a bytecode that includes \emph{swap directives} describing when to transfer data between storage and memory.
Finally, the memory program is given to \sys{}'s interpreter, which executes it using the SC protocol.

For multi-party protocols, the parties run separate instances of \sys{}'s interpreter.
In the case of garbled circuits, garbled gates are streamed from the garbler to the evaluator, as described in \secref{s:streaming}.
Both the garbler and evaluator use \sys{} to follow a memory program and run with constrained memory.

Our approach of including swap directives in the memory program relies on the planner knowing how much memory will be available at runtime.
An alternative approach is for memory programs to be agnostic to the amount of available memory.
This would add runtime overhead, as \sys{}'s interpreter would need to decide which pages to evict.
In contrast, our approach moves this overhead to the planning phase, keeping the execution phase as lightweight as possible.

\subsection{Address Translation in \sys{}}\label{s:translation}
The application programmer should not have to manage paging, so it is natural to write DSL programs in a virtual address space that is, in effect, infinitely large.
Central to designing \sys{} is deciding at which point in \figref{fig:overview} to translate this address space into a physical address space that fits in RAM.

One possibility (which \sys{} does not use) is to perform address translation at runtime, using standard operating system mechanisms for prefetching and address translation.
At runtime, swap directives in the memory program would ask the operating system to page parts of the virtual address space out to storage or in to RAM.
Unfortunately, the existing way for a Linux process to do this---the \texttt{madvise} system call---is too limited.
As of Linux 5.10, pages brought into RAM using the \texttt{MADV\_WILLNEED} hint are not mapped in the page table, so a minor page fault is incurred on the first subsequent access.
Similarly, the \texttt{MADV\_PAGEOUT} hint merely marks pages as inactive; it does not swap out pages immediately.

In contrast, \sys{} does not rely on OS address translation for demand paging.
\sys{}'s engine moves data between memory and storage via explicit I/O operations, so that its resident set size never exceeds the available RAM.
At the surface, this is similar to buffer management in a DBMS.
But unlike a DBMS, \sys{}'s planner can be viewed as solving an address translation problem in advance.
The DSL variables declared by the programmer exist in a \emph{\sys{}-virtual address space}, and the final memory program output by the planner references data (i.e., wire values) in a \emph{\sys{}-physical address space} that fits within RAM.
\sys{}'s planner creates these address spaces and performs their translation in software during the planning phase.
It includes swap directives in the memory program so that the interpreter does not run out of RAM.

To avoid confusion, we will refer to the addresses created by the OS and sent over the memory bus as \emph{OS-virtual addresses} and \emph{OS-physical addresses}.
At runtime, \sys{}'s interpreter stores the program's memory in an array, and each \sys{}-physical address in the memory program is treated as an index into this array.
Thus, \sys{}-physical addresses roughly correspond to the OS-virtual addresses of \sys{}'s interpreter.

\sys{}'s approach to address translation has several advantages.
First, in contrast to an \texttt{madvise}-based approach, \sys{}'s planner has nearly complete control over when pages are brought into memory and evicted to storage.
Second, by translating addresses in the planner, \sys{} avoids address-translation-related overheads at runtime.
In contrast, relying on OS address translation would mean minor page faults, page table updates, and TLB invalidations at runtime.

\sys{}'s approach also has a few drawbacks, however.
First, the planning phase takes longer because \sys{}'s planner must translate all addresses in software.
Second, memory programs are considerably larger because they must contain not only swap directives, but also a copy of the program translated to operate on \sys{}-physical addresses.
In particular, the memory program's length is proportional to the program's execution time because a variable local to a function or loop could be assigned different physical addresses each time the function is called or on each iteration of the loop.

Overall, we felt that the advantages of this design outweighed its drawbacks.
Longer planning times seemed reasonable because planning can happen offline and the resulting memory program can be used repeatedly.
The larger memory program size was an acceptable tradeoff because \sys{}'s planner materializes an unrolled form of the program anyway to run Belady's algorithm.
Meanwhile, \sys{}'s planner is afforded nearly full control of page eviction and replacement and \sys{}'s runtime overheads remain relatively small.

\subsection{\sys{}'s Bytecode Representation}\label{s:ir}
Recall that \sys{}'s planner expresses the program as an unrolled (branch-free) bytecode, and performs transformations on it to compute the memory program bytecode.
What operations should the bytecode instructions support?

One possibility would be for the bytecode to describe low-level operations similar to those supported by a CPU, excluding control flow instructions.
Unfortunately, such a bytecode includes the raw memory trace of the program, which, as discussed in \secref{sec:introduction}, can be impractically large.

One alternative, used by PCF~\cite{kreuter2013pcf} and Frigate~\cite{mood2016frigate}\footnote{Unlike \sys{}, these systems also include control flow operations.} (but not \sys{}), is to have each instruction correspond to a gate in the circuit $C$ being executed.
This approach would require a \emph{protocol driver} in \sys{}'s interpreter that executes each gate using the SC protocol.
To understand why this is inefficient, consider garbled circuits, for which gates are binary and wires represent bits.
The programmer specifies the circuit in terms of operations on high-level types such as integers, which are then compiled into bit-level operations.
Thus, each time the program performs a high-level operation (e.g., adding two integers), the same subcircuit (e.g., describing integer addition in terms of binary gates) is repeated in the bytecode.

To eliminate this repetition, \sys{} has each instruction describe a high-level operation directly.
This requires not only a \emph{protocol driver}, but also an \emph{engine} in \sys{}'s interpreter that expands each instruction into the relevant subcircuit at runtime.
\sys{}'s planner does not need to materialize the subcircuits because wires internal to the subcircuits are very short-lived and therefore can be ignored.

\begin{figure}[t]
    \centering
    \includegraphics[width=\linewidth]{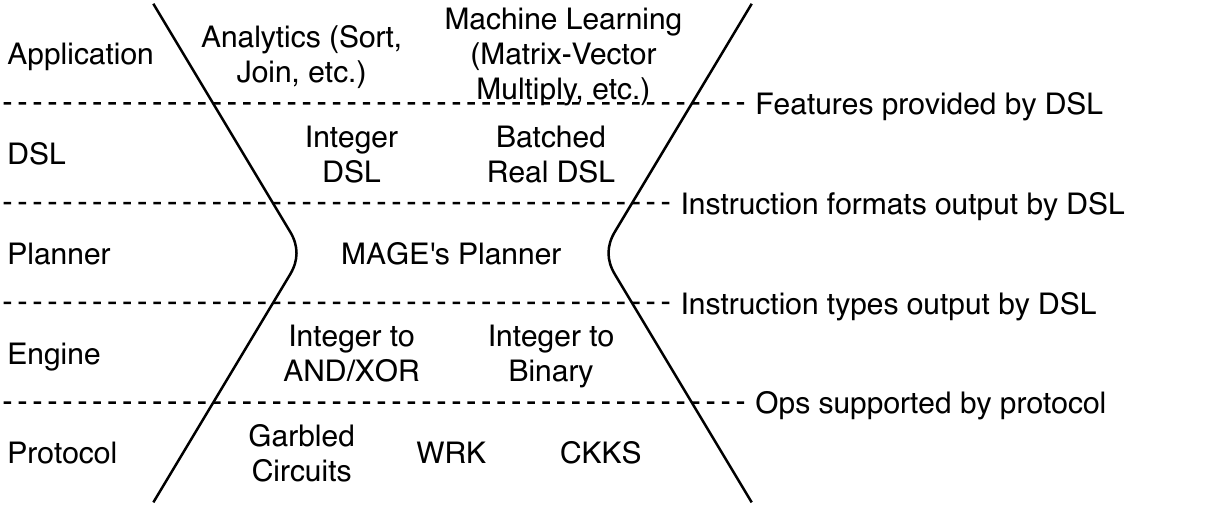}
    \caption{\sys{}'s envisioned ecosystem, with planning as the narrow waist}
    \label{fig:layering}
\end{figure}

\subsection{\sys{}'s Ecosystem and its Extensibility}\label{s:architecture}

An important consideration in \sys{}'s design is to be applicable to a range of SC protocols.
For example, garbled circuits and homomorphic encryption (CKKS) have quite different computation models, yet we show how \sys{} captures both.
\sys{}'s envisioned ecosystem can be understood as a set of layers with a narrow waist, as shown in \figref{fig:layering}.
The narrow waist is \sys{}'s planner; \sys{}'s core planning algorithms can be used with a variety of applications and interpreters.

\sys{}'s interpreter has two layers.
The upper layer, called the \emph{engine}, decomposes each instruction into a subcircuit of gates supported by the target SC protocol (\secref{s:ir}).
The lower layer, called the \emph{protocol driver}, evaluates gates with the SC protocol.
For example, when using a protocol that supports only binary AND and XOR operations (e.g., garbled circuits), one must use an engine that decomposes each instruction into a circuit of AND and XOR gates.
In contrast, when using a protocol that supports all types of binary gates (e.g., TFHE~\cite{chillotti2016faster}), one can use an engine that uses all types of gates.

One must choose compatible implementations at each layer.
For example, once one has selected an SC protocol, one should choose an engine that executes each instruction using operations supported by that protocol.
Then, one should select a DSL that outputs instructions that the chosen engine understands.
Finally, one must write the application in that DSL.

\sys{}'s planner, however, is universally compatible, allowing it to be the ``narrow waist'' of the ecosystem.
The first reason is that \sys{}'s planner does not have to understand what each instruction \emph{does}, only what memory it accesses.
Thus, even if a new instruction is introduced into a DSL, extending a header file to specify its format (which includes which fields are memory addresses) is enough for the planner to understand that instruction.
The second reason is that \sys{}'s planner does not introduce any new instructions except for swap directives, which all engines understand.
Thus, if an engine understands the instruction types output by \sys{}'s DSL, then the engine will also be able to interpret the planner's output (i.e., the memory program).

A number of frameworks and DSLs for SC~\cite{hastings2019compilers, viand2021compilers} aim to make it easier for non-SC-experts to use SC.
In contrast, \sys{} is an efficient SC execution engine; its DSLs are not necessarily geared toward non-experts, do not optimize the resulting circuit, and might expose low-level SC operations.
We discuss how these frameworks fit into \figref{fig:layering} in \secref{sec:related}.

\section{\sys{}'s Engine}

\sys{}'s execution engine is an interpreter for the final memory program.
First, it allocates an array to store the program's data.
Each \sys{}-physical address is an index into this array.
To execute an instruction, \sys{} reads the instruction's arguments from this array, makes calls to the protocol layer to compute the output, and writes the output back to the array.
Each instruction in the memory program references its input and output data directly by \sys{}-physical address; the engine sees no \sys{}-virtual addresses.
Some instructions, such as those requesting pages to be transferred between storage and memory, are handled directly by the engine, without calling the protocol.
We call such instructions \emph{directives}.

\subsection{Parallel/Distributed Engine}
SC is resource-intensive, so it is natural to scale SC by executing the protocol in a distributed fashion across \emph{multiple CPU cores} or \emph{multiple machines}.
The multiple-machine case is useful to overcome resource constraints associated with a single machine such as limited CPU cores, limited storage I/O, or, in the case of SMPC, limited network bandwidth.
This is different from having multiple parties in SMPC.
Here, we are parallelizing a single trust domain---for example, a single logical party in SMPC may execute using multiple machines.

\sys{}'s engine supports distributed execution across multiple \emph{workers}.
Each worker is a thread of computation, running \sys{}'s engine, operating on its own memory region (a \sys{}-physical address space).
Workers differ from OS processes as follows: (1) each worker contains exactly one thread, (2) workers are not necessarily isolated by hardware such as an MMU---multiple workers in a \sys{} computation could, in principle, run within the same process, and (3) memory is statically partitioned among the workers.

\sys{}'s planner does not automatically infer how to parallelize the computation.
Rather, the programmer writes DSL code in a distributed memory model, explicitly indicating asynchronous network operations to transfer data among the different workers.
The resulting memory program bytecode contains \emph{network directives} that the engine interprets.
Similarly, the protocol driver must be written to function properly when the computation is distributed over multiple workers.

Programs for \sys{} are parameterized by the Worker ID.
\sys{}'s planner is run once \emph{for each worker}.
To generate the memory program for a worker, the planner processes only the accesses for that worker---it does not need to consider other workers' accesses, because each worker can only access its own memory region.
Thus, the workers' memory programs can be generated independently and in parallel.

Using a distributed memory model provides two benefits.
First, it allows \sys{} to be agnostic to whether workers are placed on a single machine or across multiple machines.
Second, it guarantees that the access pattern for each region of memory consists of a single well-defined sequence, simplifying planning.
To ease the difficulty of explicitly specifying network transfers, one can build easier-to-use DSL libraries for common communication patterns (e.g., our implementation provides a \texttt{ShardedArray<T>} abstraction).

\begin{figure}[t]
    \centering
    \includegraphics[width=\linewidth]{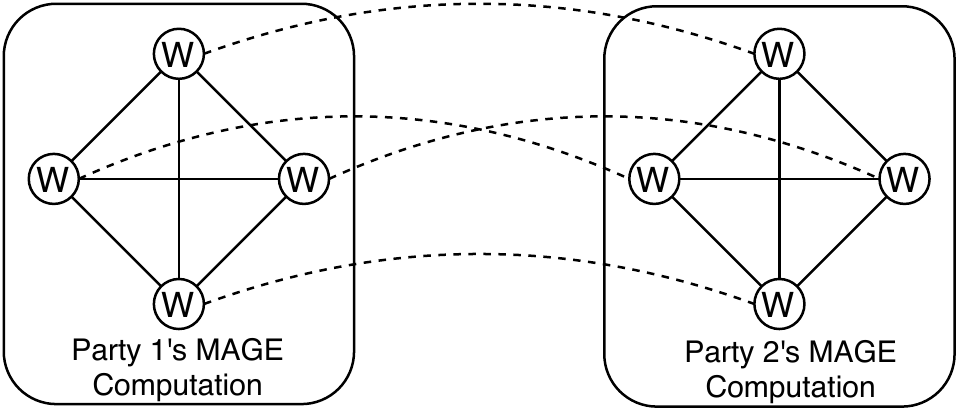}
    \caption{Example of distributed SMPC with \sys{}. Workers are denoted as circles with W. Solid lines indicate connections managed by \sys{}'s engine; dashed lines indicate connections managed by the protocol driver}
    \label{fig:networking}
\end{figure}

\subsection{Distributed SMPC}
Some SC protocols, like SMPC, require interaction over the network between mutually distrusting parties.
For such protocols, each party runs a separate \sys{} computation, with its own set of workers.
Whereas the \sys{} engine handles \emph{intra-party communication} between workers in the same party, the protocol implementation handles \emph{inter-party communication} among workers in different parties.
The inter-party topology is up to the protocol driver; our protocol driver for garbled circuits uses a one-to-one inter-party topology (\figref{fig:networking}).

\section{\sys{}'s Planner}\label{sec:design}

Our memory programming approach is to calculate the memory access pattern in advance and use it to preplan memory management.
One can potentially preplan the following:
\begin{itemize}[leftmargin=*,noitemsep,topsep=0ex]
    \item \parhead{Placement} How should we divide up a circuit into pages?
    \item \parhead{Ordering} In what order should we evaluate the gates in the SC circuit to result in the best memory behavior?
    \item \parhead{Scheduling} When should pages that will be used in the future be swapped in from storage?
    \item \parhead{Replacement} How should we choose pages to evict when making room for pages from storage?
\end{itemize}
\sys{} produces an approximate solution, using a heuristic for placement and optimizing scheduling and replacement.
Note that \sys{} does not optimize ordering; it evaluates gates in the order implicit in the DSL program for the circuit.\footnote{Optimizing ordering may be $\NP$-hard~\cite{laakeri2020algorithm}. A system that does so would be very powerful---for example, it would automatically block a loop join or tile a matrix multiplication. It is beyond the scope of this work.}

\subsection{Organization of \sys{}'s Planner}\label{s:planner_organization}

\begin{figure}[t]
    \centering
    \includegraphics[width=\linewidth]{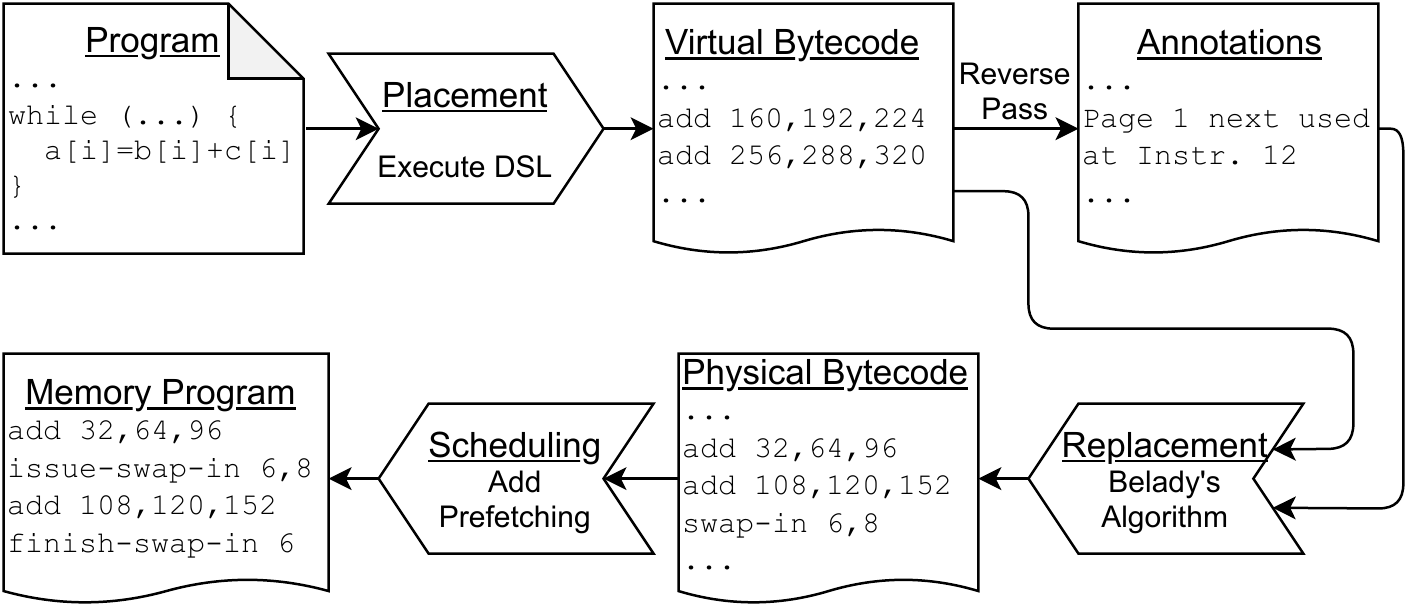}
    \caption{\sys{}'s planner's workflow, with its three stages}
    \label{fig:planner}
\end{figure}

We organize \sys{}'s planner into stages (\figref{fig:planner}):
\begin{enumerate}[leftmargin=*,topsep=0ex,noitemsep]
    \item \textbf{Placement.} This stage accepts a DSL program and organizes wires into \sys{}-virtual pages.
    It outputs instructions referencing wires by \sys{}-virtual address. 

    \item \textbf{Replacement.} This stage adds instructions to swap pages to/from storage, deciding which pages to evict.
    It outputs instructions referencing wires by \sys{}-physical address.

    \item \textbf{Scheduling.} This stage moves swap instructions within the instruction stream and relocates wires to mask the latency of moving data between memory and storage.
\end{enumerate}

For a parallel/distributed program, \sys{}'s planner is invoked separately for each worker, with separate \sys{}-virtual and \sys{}-physical address spaces.
Network directives in the program transfer data among those address spaces.

\sys{}'s planner does not benefit from \sys{}'s memory programming techniques, so it is important that planning does not consume an unreasonable amount of memory.
We keep the planner's memory usage lightweight by (1) writing/reading the intermediate bytecodes to/from files instead of keeping it all in memory, (2) designing the DSLs to be lightweight, and (3) keeping track of pages instead of individual bytes.

\subsection{\sys{}'s First Stage: Placement}
\sys{}'s placement module is, in effect, a page-aware memory allocator for the DSL.
It unrolls the DSL, allocating space for each variable and intermediate value in the \sys{}-virtual address space.
It outputs a bytecode for the program in which each variable is referenced by its \sys{}-virtual address.

\subsubsection{Unrolling the DSL Code}\label{s:unrolling}
\sys{}'s DSLs are internal to C++.
This means that the DSL is a set of convenient C++ APIs to specify the program's behavior, often involving operator overloading.
The program is specified as a C++ function that uses these APIs.

\figref{fig:dsl} shows a program that solves Yao's Millionaire's problem~\cite{yao1982protocols}.
\texttt{Integer<width>} describes an Integer datum with the specified width in bits.
\texttt{Bit} is an alias for \texttt{Integer<1>}.

\sys{}'s planner does not parse the DSL program's source code or manipulate its AST.
Instead, it simply calls the C++ function containing the DSL program.
As the DSL code executes, it produces a bytecode describing the computation.
For example, the overloaded \texttt{+} operator for \texttt{Integer} emits an \texttt{Add} instruction in the output bytecode; it does not actually add integers using secure computation.
Each output instruction references its operands by \sys{}-virtual address.
Thus, the DSL (e.g., the \texttt{Integer} class) calls \sys{}'s placement module to allocate memory in the \sys{}-virtual address space for intermediate results, including those stored in variables.

For example, see \figref{fig:dsl}.
On the \texttt{mark\_input} and \texttt{>=} operations, an allocation request is made to \sys{}'s placement module to obtain a \sys{}-virtual address, and an instruction is emitted to perform that operation (obtain input or integer comparison) and store the result at that \sys{}-virtual address.
Once an \texttt{Integer}'s destructor is called, or if an \texttt{Integer} is reassigned to a new \sys{}-virtual address, a deallocation request is made to \sys{}'s placement module for the \sys{}-virtual address previously held by that \texttt{Integer}.

For a parallel/distributed program, the worker ID and total number of workers are provided via the \texttt{ProgramOptions} structure.
The C++ code can branch on these variables, to have each worker operate differently and exchange data appropriately to perform the parallel/distributed computation.

Each \texttt{Integer} object contains only the \sys{}-virtual address of its contents; other attributes, such as width, are template arguments and do not consume memory.
Thus, \texttt{Integer}s and other DSL-provided data types are typically smaller than the encrypted data items they represent.
For example, a 32-bit integer encrypted for the garbled circuit protocol is 1~KiB in size, whereas an \texttt{Integer<32>} object used during planning is just 8~B (a single \sys{}-virtual pointer).
This helps keep the memory cost of the planning phase small.

\begin{figure}[t]
    \centering
\begin{lstlisting}[language=C++, columns=flexible, basicstyle=\ttfamily]
void millionaire(const ProgramOptions& args) {
    Integer<32> alice_wealth, bob_wealth;
    alice_wealth.mark_input(Party::Garbler);
    bob_wealth.mark_input(Party::Evaluator);
    Bit result = alice_wealth >= bob_wealth;
    result.mark_output();
}
\end{lstlisting}
    \caption{Example code in an Integer-based DSL internal to C++ to solve Yao's Millionaire's problem}
    \label{fig:dsl}
\end{figure}

\subsubsection{Memory Allocation Strategy}

When \sys{}'s placement module allocates memory for a variable, it ensures that the variable is contained in a single \sys{}-virtual page; a variable must never straddle two pages.
The reason is that two adjacent \sys{}-virtual pages may not be adjacent in the OS-virtual address space at runtime.

A key issue in designing the placement module's memory allocator is internal fragmentation~\cite{randell1969fragmentation, denning1970virtualmemory}.
Some fragmentation, which we call \emph{classic fragmentation}, arises from the inability to pack variables onto pages (e.g., part of a page's space cannot store any variable).
Another type of fragmentation, which we call \emph{effective fragmentation}, arises from the page's lifetime exceeding some of the variables it stores; if even one wire on a page is alive, the entire page remains alive.

To reduce classic fragmentation, \sys{}'s placement stage uses techniques from slab allocators~\cite{bonwick1994slab}.
Each page contains only variables of a particular size.
When a variable goes out of scope in the DSL, its ``slot'' in its page is marked as free.
When a space for a variable must be allocated, \sys{}'s placement module looks for a free slot in a page containing variables of that size; if no such pages have free slots, it allocates a new page for variables of that size.
The slab size is one \sys{}-virtual page.
This ensures that no variable will straddle a page boundary.
Just as in slab allocators, some leftover space at the end of a page may be unusable, but this can be controlled by tuning the page size.
Unlike slab allocators, \sys{}'s placement module does not preserve object state across allocations.

To reduce effective fragmentation, \sys{}'s placement stage uses the following heuristic when allocating memory for a variable.
If multiple pages, for the specified variable size, have free slots available, then \sys{} uses the candidate page with the fewest free slots.
This allows the number of live pages to decrease if the number of live variables decreases, by giving a chance for all variables on a page to die.

\subsection{\sys{}'s Second Stage: Replacement}\label{s:replacement}
We apply Belady's MIN algorithm~\cite{belady1966study}.
MIN is theoretically optimal in the number of \swapin{} operations, but it does not minimize the number of swap operations if \swapout{} operations are also considered.
The reason is that only dirty pages need to be written back to storage (i.e., ``swapped out'').
Minimizing the number of swaps when taking this into account is $\NP$-hard~\cite{farach1998lra}.
Regardless, MIN produces a solution with at most $2\times$ as many swaps as the theoretical optimum,\footnote{This occurs in the worst case where it evicts only dirty pages, but there is an optimal solution that evicts the same number of clean pages.} so it is useful in \sys{}'s replacement stage.

To use MIN, we first make a backward pass over the program to determine, each time a page is used, the time (instruction ID) at which it is used next.
Then we make a forward pass over the program, using the annotated next use time to determine which page to swap out.
This requires us to maintain a priority queue of resident pages, so that we can quickly identify which one's next use is farthest in the future.
Each instruction, even if its arguments are already resident, requires us to also perform a \texttt{decrease\_key} operation on the priority queue to adjust pages' next use time.
Therefore, if $N$ is the number of instructions and $T$ is the number of pages that fit in memory, applying Belady's MIN algorithm is $\bigO(N\log T)$.

This stage outputs an instruction stream that contains swap directives and references wires by \sys{}-physical address.
To support this, \sys{}'s planner maintains a data structure that maps \sys{}-virtual page numbers to \sys{}-physical frame numbers, similar to a page table.

When planning a parallel/distributed program, the planner must be careful to not steal a page that is currently being used for network I/O.
Thus, \sys{}'s replacement phase reads the network directives to infer the outstanding asynchronous network operations.
When stealing pages, it issues \emph{network barrier} directives, as necessary, to ensure that the engine waits for the relevant network I/Os to complete.

\subsection{\sys{}'s Third Stage: Scheduling}
We introduce a parameter $\ell$ called the \emph{lookahead}.
To prefetch data, \sys{}'s scheduling algorithm attempts to move \swapin{} directives $\ell$ instructions earlier in the instruction stream.
However, this does not work if one of the $\ell$ intervening instructions uses the page frame into which we are bringing in data.
We solve this by budgeting $B$ extra physical page frames, called the \emph{prefetch buffer}; the replacement stage is now run with a capacity of $T - B$ frames, not $T$ frames.
Data is brought asynchronously into a free slot in the prefetch buffer.
Only when it is finally needed is it copied from the prefetch buffer into its destination physical page frame.
Instead of \swapin{} directives, the memory program contains \iswapin{} directives, which initiate the transfer of a page into memory, and \fswapin{} directives, which block execution until a swap operation has completed.
Ideally, swap operations will be scheduled such that \fswapin{} never blocks, but it serves as an important fallback to prevent old/corrupt data from being used if the transfer is unpredictably delayed.

We use the prefetch buffer similarly to swap out pages.
The page to be swapped out is copied into a free slot in the prefetch buffer and then swapped out to storage with an \iswapout{} directive while execution of subsequent instructions continues.
Unlike \swapin{} operations, there is no clear deadline by which the write to storage must complete.
Thus, we delay issuing a \fswapout{} directive for as long as possible; we only issue it when allocating a slot in the prefetch buffer fails.
In such a situation, we identify the oldest \iswapout{} operation, issue the \fswapout{} directive for it, and reclaim its page in the prefetch buffer.

One could eliminate the copying of pages to/from the prefetch buffer by rewriting future instructions.
We did not implement this optimization because it would introduce additional complexity and \sys{} performs well without it.

A natural question is how large $B$ must be.
SSDs have bandwidths less than 10 GB/s and latencies that are usually less than 1 ms.
Based on these measurements, Little's Law gives: $B = 10\text{ GB/s} \cdot 1\text{ ms} = 10\text{ MB}$.
For server-class machines, this is $< 1$\% of physical memory.
In practice, we use $16$--$32\text{ MiB}$ to account for burstiness/queuing, still only a small fraction of available memory.
Thus, \sys{}'s scheduling promises to mask storage latency with only a small memory penalty.

\section{Implementation}\label{sec:implementation}
We implemented a prototype of \sys{} in C++, including support for two protocols: garbled circuits and CKKS.
Using \texttt{cloc}, we found that our implementation is $\approx 11,000$ lines of code, excluding comments and blank lines, broken down as follows:
$\approx 2,800$ for common libraries used throughout \sys{} (e.g., data buffering for I/O, configuration file parsing, etc.);
$\approx 1,300$ for \sys{}'s planner;
$\approx 900$ for protocol drivers (not including the underlying cryptography);
$\approx 1,000$ for \sys{}'s DSLs and libraries for those DSLs (e.g., for sharding data);
$\approx 1,100$ for \sys{}'s engines;
$\approx 1,600$ for SC programs written in \sys{}'s DSLs, used for testing and evaluating \sys{};
$\approx 1,900$ for the underlying cryptography for garbled circuits, much of which is based on EMP-toolkit~\cite{emp-toolkit};
and $\approx 400$ for in-progress (not yet complete) support for a third protocol.
We build \sys{} using \texttt{clang++} version 10.0.0 with the optimization flags \texttt{-Ofast -march=native}.
\sys{} runs as a Linux process, with no changes to kernel code.

\subsection{\sys{}'s Interpreter}\label{s:interpreter}
\parhead{Engine} The \texttt{Engine} class implements common functionality for the engine layer, including support for directives.
It establishes pairwise TCP connections among workers within a single party, to support network directives.
Swap directives are implemented using the \texttt{aio} facility provided by the kernel (not to be confused with POSIX \texttt{aio}); the swap file/device is \texttt{open}ed with the \texttt{O\_DIRECT} flag.
\sys{} engines are implemented as class templates that extend (inherit from) the \texttt{Engine} class.
The protocol driver class is provided to the engine as a template argument, so the engine can make calls to it.
We avoided using virtual functions for this, as their overhead can be significant (e.g., for free XORs).

\parhead{Protocol Driver} The protocol driver exposes the SC protocol's native operations to the engine as a set of methods.
When the engine invokes these methods, it provides pointers to data to operate on, stored in a large array representing the \sys{}-physical address space.
The protocol driver specifies the type of entries in the engine's array, in effect dictating what each \sys{}-physical address actually corresponds to for its protocol (plaintext bits, ciphertext bytes, etc.), and provides a plugin to the DSL so it can allocate \sys{}-virtual memory accordingly.
The protocol driver must not store pointers to dynamically allocated memory in the array.
The reason is that the engine swaps out only the contents of the array, not including any dynamically-allocated memory it points to.
In addition to the SC protocol's cryptographic routines, the driver manages all protocol-specific operations.
This includes sharing protocol-specific state among workers within a party, obtaining input data, writing output data, and managing intra-party communication where necessary (e.g., sending garbled gates from the garbler to the evaluator).

\subsection{Extending \sys{} with New Protocols}
To extend \sys{} with a new protocol, one must, at minimum, write a protocol driver to support it.
If the operations exposed by the new protocol driver are identical to those exposed by an existing protocol driver, then one can use the same engine that works with the existing protocol.
Otherwise, one must implement a new engine or modify an existing engine.
This involves deciding which instruction types the new engine will be compatible with.
If the supported instruction types differ from what existing DSLs produce, then one may have to implement a new DSL or modify an existing DSL.

We implemented protocol drivers for garbled circuits and CKKS.
Garbled circuits and CKKS support different operations, so we implemented a separate DSL (Integers vs. Batches) and engine (AND-XOR vs. Add-Multiply) for each protocol.
This conveniently allows us to showcase \sys{}'s ability to support different implementations of each layer.
That said, it is not uncommon for related SC protocols to expose similar interfaces.
For example, the WRK protocol~\cite{wang2017ag2pc, wang2017agmpc} exposes the same interface as garbled circuits (AND-XOR), so support for WRK, if added, could reuse our Integer DSL and AND-XOR engine.

\subsection{Garbled Circuit Protocol Driver}
For garbled circuits, wires have uniform size, so we allow \sys{} address spaces to be wire-addressed; the DSL is unaware of the size of wires in bytes.
Some subcircuits used by the AND-XOR engine are based on those used by Obliv-C~\cite{zahur2015oblivc}.
Our garbled circuit driver uses cryptographic kernels from EMP-toolkit~\cite{emp-toolkit}.
We implement oblivious transfer (OT) using multiple background threads.
Concurrently with our work, EMP-toolkit was updated to use the MiTCCRH hash function~\cite{guo2019mitccrh}; our implementation is based on an older version of EMP-toolkit based on fixed-key AES~\cite{bellare2013fixedkey}.
When we compare \sys{} to EMP-toolkit in \secref{sec:evaluation}, we use the older version of EMP-toolkit so the comparison is fair.
This is not a limitation of \sys{}; our driver could be changed to use MiTCCRH.

\subsection{CKKS Protocol Driver}
CKKS ciphertexts vary in size depending on their level, so for CKKS' DSL and engine, \sys{} address spaces are byte-addressed.
The protocol driver provides a plugin to the DSL describing the particular wire sizes in bytes.
It uses the CKKS implementation in Microsoft SEAL~\cite{sealcrypto}.
We chose parameters for CKKS that allow a multiplicative depth of 2.
A challenge was that SEAL ciphertext objects contain pointers and dynamically-allocated memory.
\sys{} cannot swap such objects to storage (see \secref{s:interpreter}).
Thus, TE protocol driver serializes ciphertexts using SEAL's built-in serialization methods when they are not in use; each operation (e.g., add, multiply) deserializes the arguments, computes the result, and then serializes the result.
We quantify the cost of serialization in \secref{sec:evaluation}.
This overhead is not fundamental; CKKS ciphertexts could be implemented as flat buffers, or homomorphic operations could be implemented to operate directly on serialized ciphertexts.

After a multiplication, CKKS ciphertexts are typically relinearized and rescaled before the next multiplication.
But if two products are added (e.g., $ab + cd$), one can perform relinearization once for the overall result instead of for each multiplication separately (e.g., $ab$ and $cd$).
\sys{}'s DSL supports this optimization, which is crucial to achieve good performance on \rstats{} and the linear algebra workloads.

\section{Evaluation}\label{sec:evaluation}
\subsection{Workloads}\label{s:workloads}
We now establish a set of SC workloads for our evaluation.
Garbled circuits and CKKS support different operations---bitwise operations for garbled circuits, and add-multiply circuits of low multiplicative depth for CKKS---so we design separate workloads for each protocol.
These workloads are data-intensive ``kernels'' that may be used as part of larger SC applications.
We discuss larger SC applications in \secref{s:applications}.

\subsubsection{SMPC Collaborative Applications}
One application of SMPC is federated data analytics~\cite{volgushev2019conclave, poddar2021senate}.
Aggregations (\texttt{GROUP BY} operations) and joins are particularly memory-intensive.
A federated data analytics system may express equi-joins as set intersections (SI) and aggregations as set unions (SU), both of which can be implemented by merging sorted lists~\cite{poddar2021senate}.
This inspires our first benchmark, \merge{}: \emph{merging sorted lists of records}.
In some cases, the input lists may not be already sorted.
This inspires our second benchmark, \sort{}: \emph{sorting a list of records}.
For joins other than equi-joins, the system must fall back to a classic loop join.
This is our third benchmark, \ljoin{}: \emph{loop join}.
For concreteness, we assume that each record is 128 bits long, and that the first 32 bits are the key used for sorting or joining; the problem size $n$ is the number of records per party.

Privacy-preserving machine learning applications inspire our fourth benchmark, \mvmul{}: \emph{matrix-vector multiply with 8-bit integers}.
A recent proposal for secure neural network inference, \textsc{Xonn}~\cite{riazi2019xonn}, suggests \emph{binarizing} the neural network.
This inspires our fifth benchmark, \binfclayer{}: \emph{binary fully-connected layer}.
It consists of a series of XNOR and PopCount operations similar to multiplying a binary matrix by a binary vector, followed by a binary activation function.
For simplicity, we do not include batch normalization.

\subsubsection{CKKS Homomorphic Encryption}
We restrict ourselves to workloads for which CKKS is efficient---workloads that can be expressed as arithmetic circuits of low multiplicative depth.
The sixth workload is \rsum{}: \emph{sum of a list of real numbers}, which requires no multiplications.
The seventh workload is \rstats{}: \emph{computing the mean and variance of real numbers}, which requires a multiplicative depth of 2.
These represent simple data analytics workloads; the problem size $n$ is the number of elements.

Our remaining workloads are inspired by machine learning and linear algebra.
The eighth workload is \rmvmul{}: \emph{matrix-vector multiply with real numbers}.
Finally, we consider two variants of matrix multiplication.
The ninth workload is \naivermatmul{}: \emph{matrix-matrix multiply with a na\"ive nested \texttt{for} loop}.
The tenth workload is \tiledrmatmul{}: \emph{tiled matrix-matrix multiply}.
The problem size $n$ is the length of one side of the matrix (also for \mvmul{} and \binfclayer{}).

\subsubsection{Implementation of Workloads}
For simplicity, our implementations of some of these workloads only support power-of-two sizes and power-of-two number of workers, but this is not a fundamental limitation of \sys{}.
Some workloads can, in principle, be optimized through streaming.
For example, \rsum{} could read each input one at a time, add the result to an accumulator, and then output the accumulator, instead of holding the entire input dataset in memory.
We deliberately avoided such ``optimizations,'' as they would not be possible if the workload were part of a larger computation whose intermediate results are held in memory.
Thus, each workload operates in three non-overlapping phases: (1) the inputs are read into memory, (2) the computation is performed, materializing the output in memory, and (3) the output is written to a file.

For the parameters we chose, the CKKS scheme encrypts vectors of dimension 4096.
Thus, each of our workloads for CKKS could be applied to 4096 instances of the problem in a SIMD fashion with no additional overhead.
There are ways to use the 4096 slots in the vector to speed up a \emph{single} problem, for example, by vectorizing matrix multiplication~\cite{jiang2018secure}.
Our workloads, for simplicity, do not apply such techniques, but \sys{} is not incompatible with them.

\subsection{Empirical Methodology}\label{s:scenarios}
We compare \sys{}'s performance to an upper bound and a lower bound.
The upper bound, \textit{OS Swapping}, is the speed when relying on the operating system's paging.
The lower bound, \textit{Unbounded}, is the speed when the entire computation fits in memory.
We measure these three scenarios as follows:

\newcommand{\mysolution}[1]{\noindent\textit{#1}}
\mysolution{1. Unbounded.}
\sys{}'s planner is run assuming enough memory to fit the program.
Thus, \sys{}'s planner does not insert swap directives in the memory program.
Finally, \sys{}'s engine executes the memory program outside of any \verb|cgroup|.

\mysolution{2. OS Swapping.}
A memory program is generated in the same way as for the \emph{Unbounded} solution.
However, it is executed in a \verb|cgroup| that limits physical memory to a fixed amount.

\mysolution{3. \sys{}.}
\sys{}'s planner is run assuming a fixed physical memory capacity, minus the prefetch buffer and the interpreter's overhead.
The resulting plan is run within a \verb|cgroup| that limits physical memory to 1 GiB or 16 GiB, to ensure that the memory overhead fits in the limit.

Except where stated otherwise, we used \texttt{D16d\_v4} instances on Microsoft Azure~\cite{azureddv4}.
We chose this instance type for a few reasons.
First, it has enough memory to fit the entire computation for most experiments, necessary for the Unbounded scenario.
Second, it contains a local ``temporary'' SSD.
We use it for swap space (one of its recommended uses~\cite{azuretempdisk}) and for the file containing the memory program.
Third, it provides enough network bandwidth so as not to be a bottleneck for garbled circuits (we explore the WAN setting in \secref{s:wan}).

We set \sys{}'s parameters as follows.
For garbled circuits, we used a page size of $64$ KiB, lookahead $\ell$ of 10,000 instructions, and prefetch buffer size $B$ of 256 pages.
For CKKS, we used a page size of $2$ MiB, lookahead $\ell$ of 100 instructions, and a prefetch buffer size $B$ of 16 pages.
Because CKKS ciphertexts are large, we used a larger page size (slab size) than for garbled circuits to reduce external fragmentation.
Additionally, we left an additional 32--64 MiB of memory unused, to accommodate the memory used by \sys{}'s interpreter.

\subsection{Comparison to Existing Frameworks}
We compare \sys{}'s garbled circuits performance to that of EMP-toolkit.
Our goal is to demonstrate that \sys{}'s techniques do not limit the performance of garbled circuits compared to an existing system.
We use \merge{} for the comparison.
We implemented \merge{} in EMP-toolkit's DSL, and used EMP-toolkit's library for merging sorted arrays.

We discovered that EMP-toolkit is an order of magnitude slower than \sys{}.
This was because EMP-toolkit performs a separate invocation of OT extension, which involves a network round-trip, each time an Integer input is read for the evaluator.
Our garbled circuits implementation for \sys{} does not have this problem because it performs OTs in larger batches using background threads, regardless of the units by which the program reads the input.
To eliminate this effect, we exclude the time to read the input, for both EMP-toolkit and \sys{}, for this experiment only; we measured the time to merge the two arrays once they are materialized in memory.

We also compare \sys{}'s CKKS performance on \rstats{} to a C++ program that uses SEAL directly.
The main source of overhead in \sys{} is the need to deserialize the input ciphertexts and serialize the output ciphertext, for each instruction.

\begin{figure}[t]
    \centering
    \includegraphics[width=0.47\linewidth]{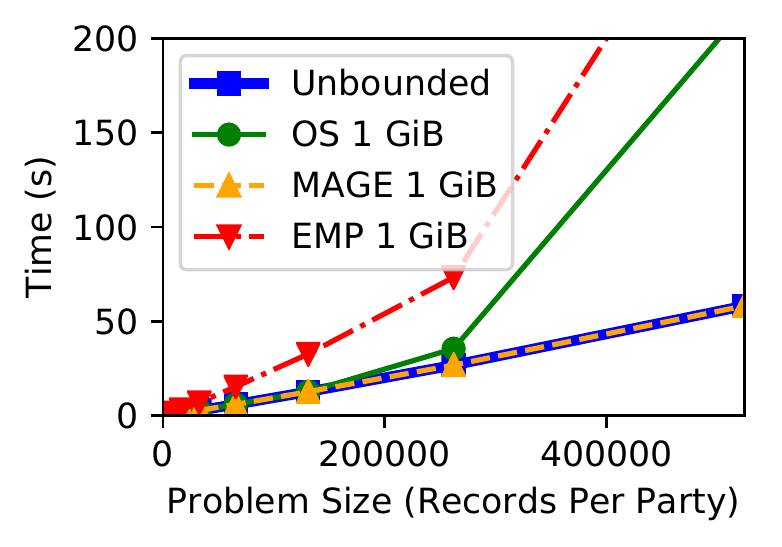}
    \hspace{-1ex}
    \includegraphics[width=0.51\linewidth]{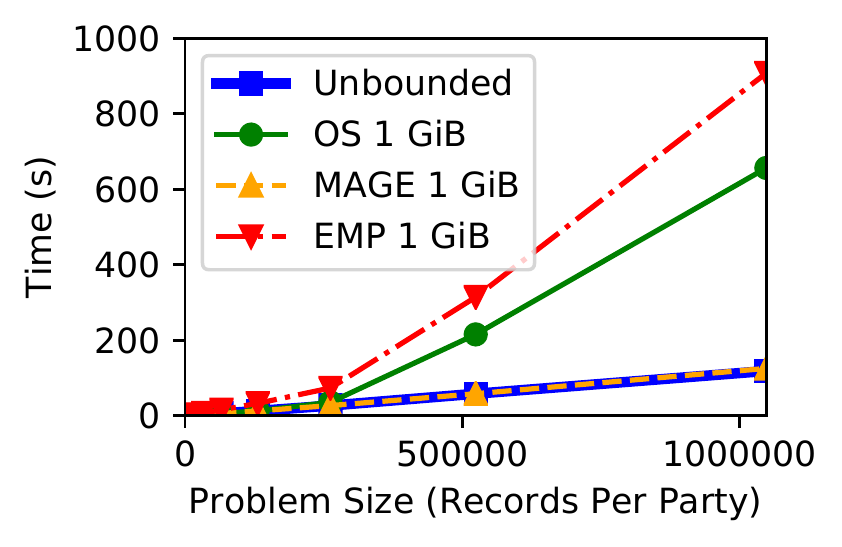}
    \caption{Comparison of \sys{} and EMP-toolkit}
    \vspace*{-0.5ex}
    \label{fig:emp_comparison}
\end{figure}

\begin{figure}[t]
    \centering
    \includegraphics[width=0.47\linewidth]{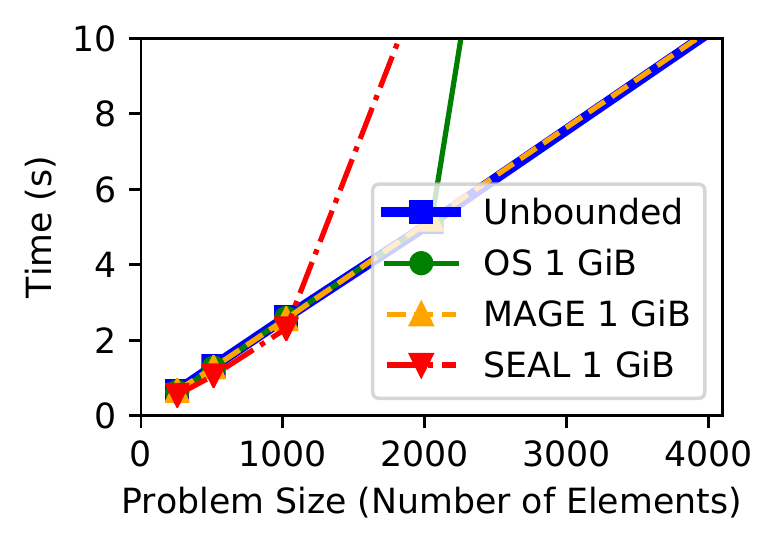}
    \hspace{-1ex}
    \includegraphics[width=0.48\linewidth]{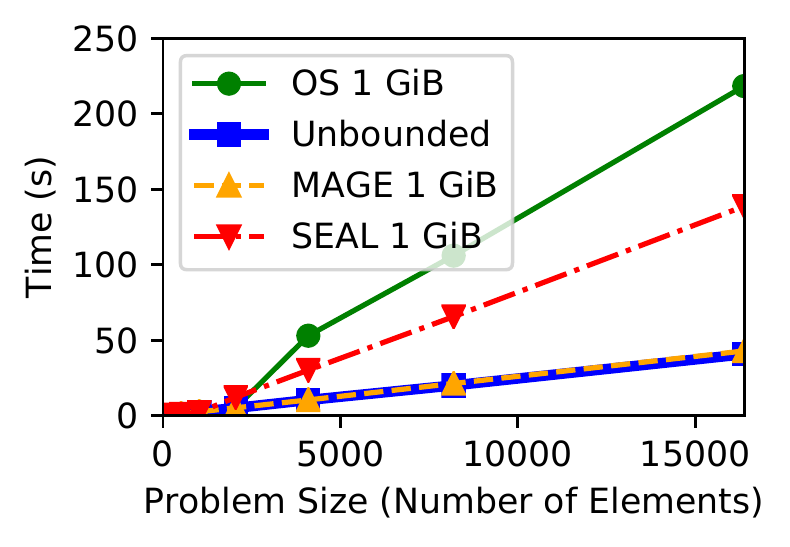}
    \caption{Comparison of \sys{} and SEAL}
    \label{fig:seal_comparison}
\end{figure}

The results are shown in \figref{fig:emp_comparison} and \figref{fig:seal_comparison}.
The graphs on the left are zoomed in to smaller problem sizes to show the point where memory demand exceeds available physical memory.
``OS'' refers to scenario 2 in \secref{s:scenarios}; ``EMP'' and ``SEAL'' refer to those systems similarly running in a \texttt{cgroup}.
EMP performs about $3\times$ worse than OS when the problem fits in memory; when it does not, the relative overhead is small ($\approx$ 33\%).
We found that EMP performs worse than OS primarily due to (1) the overhead of its ``real-time circuit optimization'' feature, (2) inefficient data buffering when using the network, and (3) virtual function overhead when executing the circuit.
OS uses \sys{}'s runtime, so it does not have these issues.
SEAL is faster than OS when the problem fits in memory, but only slightly (less than 20\%), indicating that the serialization overhead is not large.
When the problem size does not fit in memory, SEAL improves further compared to OS, but remains less than $2\times$ faster than OS.


\subsection{Overhead of Swapping Pages}\label{s:mage_overhead}
\begin{figure*}[t]
    \centering
    \includegraphics[width=\linewidth]{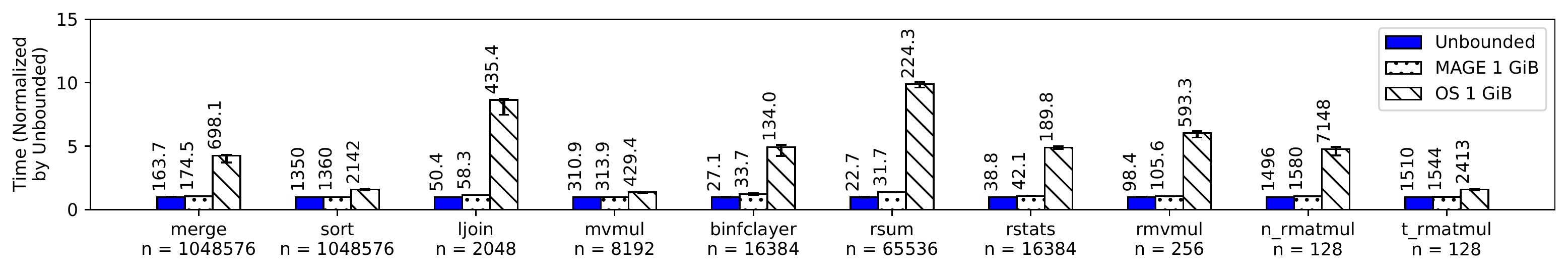}
    \caption{Performance of Unbounded, OS Swapping, and \sys{}, normalized by the time for Unbounded; absolute times, in seconds, are printed at the upper left corner of each bar}
    \label{fig:workloads_single}
\end{figure*}

\begin{figure*}[t]
    \centering
    \includegraphics[width=\linewidth]{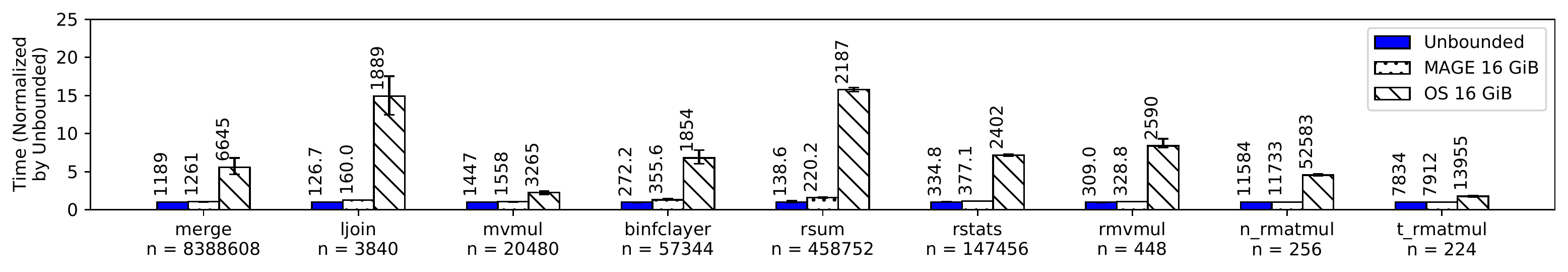}
    \caption{Repeat of \figref{fig:workloads_single}, with larger problem sizes and a 16 GiB memory limit (note the larger y-axis scale)}
    \label{fig:workloads_single_16gib}
\end{figure*}

We ran the three scenarios on all 10 workloads, using a 1 GiB memory limit.
The results are shown in \figref{fig:workloads_single}.
We ran 8 trials on different Azure instances (8 different pairs of instances, for garbled circuits) and plot the median; error bars are the quartiles.
We additionally ran experiments using a 16 GiB memory limit.
We increased the problem sizes so that their memory use exceeded 16 GiB (necessary for the OS scenario) but fit within the 64 GiB available on the virtual machines (necessary for the Unbounded scenario).
Our methodology is the same as for the 1 GiB memory limit.
We do not include \sort{} in our results for the 16 GiB memory limit, because the intermediate bytecodes produced while planning were too large for the local SSD.
The results are shown in \figref{fig:workloads_single_16gib}.
\sys{} outperforms OS swapping by at least $4\times$ on 7 of the workloads, with improvements of $\approx 12\times$ for \ljoin{} and $\approx 10\times$ for \rsum{}.
Its performance is within 15\% of Unbounded for 7 of the workloads (including \sort{} from \figref{fig:workloads_single}).

\sys{}'s improvement compared to OS is higher for \binfclayer{} and \rmvmul{} than for \mvmul{}; although all three have similar access patterns, \mvmul{} has lower memory intensity because multiplying integers in a garbled circuit has high overhead.
For complex access patterns, like \merge{} and \sort{}, \sys{}'s improvement is not markedly higher than for simple scans like \ljoin{}, \rsum{}, and \rstats{} (note that both input tables for \ljoin{} fit in memory; it is the \emph{output}, populated in order, that does not fit).
\sys{} is less affected by high memory intensity than OS, allowing it to perform well.

\begin{figure*}[t]
    \centering
    \includegraphics[width=\linewidth]{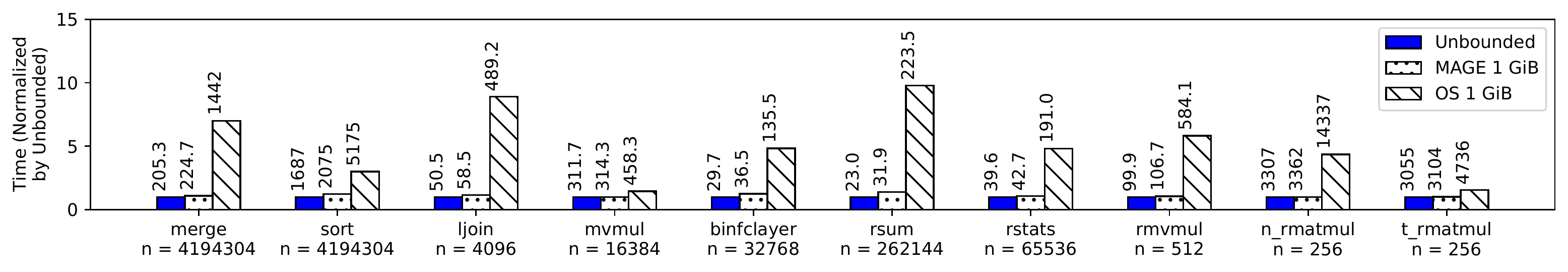}
    \caption{Normalized performance of Unbounded, OS Swapping, and \sys{}, parallelized over $p = 4$ workers (per party)}
    \label{fig:workloads_parallel}
\end{figure*}

\subsection{Overhead of Planning}
The time and peak memory use for planning each workload for the \sys{} scenario in \figref{fig:workloads_single} and \figref{fig:workloads_single_16gib} is shown in \tabref{table:planning_single}.
\iffull
These measurements were collected by running \texttt{/usr/bin/time -v} when invoking \sys{}'s planner.
\fi
Note that \sys{}'s planning is outside of the critical path: for a given circuit, \sys{}'s planner can be run before the parties' inputs are known.
For garbled circuits, although the garbled circuit \gc{} cannot be reused if the computation is re-run, \sys{}'s memory program \emph{can} be safely reused.

The planning time and final memory program size are linear in the size of the \emph{computation} (size of $C$), not in the size of the memory demand.
Nevertheless, the planning times are generally less than the time to perform the execution and the planner's memory consumption is significantly smaller than the available memory at runtime for all experiments.

Generating memory programs for CKKS is more efficient than for garbled circuits.
This is because each instruction for CKKS operates on more memory than for garbled circuits, which means that the problem sizes that fill a given physical memory size tend to require smaller bytecodes for CKKS than for garbled circuits.
For example, an instruction operating on integers in a garbled circuit program may operate on a few kilobytes of memory (each bit of each integer is 16 bytes), but for CKKS, each instruction operates on a \emph{vector} of real numbers, whose encrypted size is hundreds of kilobytes.

For CKKS, the final memory programs were $<100$~MiB for \figref{fig:workloads_single} and $<1$~GiB for \figref{fig:workloads_single_16gib}.
For garbled circuits other than \sort{}, they were $<5$~GiB for \figref{fig:workloads_single} and $<65$~GiB for \figref{fig:workloads_single_16gib}.
For \sort{}, it was less than $<25$~GiB for \figref{fig:workloads_single}. 
\sys{}'s planner requires about $4$--$5\times$ times more storage space than the final memory program due to the need to materialize intermediate bytecodes of similar size, but this could be optimized by pipelining stages of \sys{}'s planner where it is possible to do so (e.g., replacement and scheduling in \figref{fig:planner}).

\begin{table}
    \centering
    \setlength\tabcolsep{4pt}
    \begin{tabular}{|c||c|c||c|c|}\hline
        Problem & Time (\ref{fig:workloads_single}) & Mem. (\ref{fig:workloads_single}) & Time (\ref{fig:workloads_single_16gib}) & Mem. (\ref{fig:workloads_single_16gib})\\\hline\hline
        \merge{} & 38.0 & 42.6 & 291.6 & 299.4\\\hline
        \sort{} & 367.3 & 42.7 & N/A & N/A\\\hline
        \ljoin{} & 6.7 & 121.0 & 23.6 & 411.4\\\hline
        \mvmul{} & 56.0 & 527.5 & 298.2 & 3268\\\hline
        \binfclayer{} & 77.2 & 19.1 & 1041 & 165.7\\\hline
        \rsum{} & 0.04 & 9.6 & 0.29 & 30.2\\\hline
        \rstats{} & 0.04 & 10.9 & 0.34 & 48.5\\\hline
        \rmvmul{} & 0.09 & 16.4 & 0.24 & 36.9\\\hline
        \naivermatmul{} & 2.2 & 246.1 & 18.6 & 1927\\\hline
        \tiledrmatmul{} & 2.3 & 246.5 & 12.9 & 1246\\\hline
    \end{tabular}
    \caption{Planning times (s) and peak memory use of the planner (MiB) for workloads in \figref{fig:workloads_single} and \figref{fig:workloads_single_16gib}}
    \label{table:planning_single}
\end{table}

\subsection{Impact of Parallelism}
We now explore how the relative performance of Unbounded, OS, and \sys{} are affected by parallelizing the computation.
We did experiments parallelizing the computation across four workers (per party, for garbled circuits).
We place each worker on a separate VM instance, each with a separate SSD.

We ran each experiment three times, using the same cluster of machines for all trials, and report the median in \figref{fig:workloads_parallel}.
Most experiments follow a similar pattern as \figref{fig:workloads_single}, indicating that \sys{}'s performance gains persist when we parallelize the computation.
For two experiments, \merge{} and \sort{}, \sys{}'s improvement over OS Swapping visibly increases.
Whereas the other workloads are parallelized by splitting the input among the workers in a communication phase at the beginning and then computing independently thereafter, \merge{} and \sort{} have a communication phase in the \emph{middle} of the computation (several such phases in the case of \sort{}).
That OS Swapping performs worse for these workloads, but \sys{} does not, suggests that the OS virtual memory system might be introducing jitter, which interacts poorly with the communication phase and induces stragglers.

\begin{figure}[t]
    \centering
    \begin{subfigure}[t]{0.495\linewidth}
        \centering
        \includegraphics[width=\linewidth]{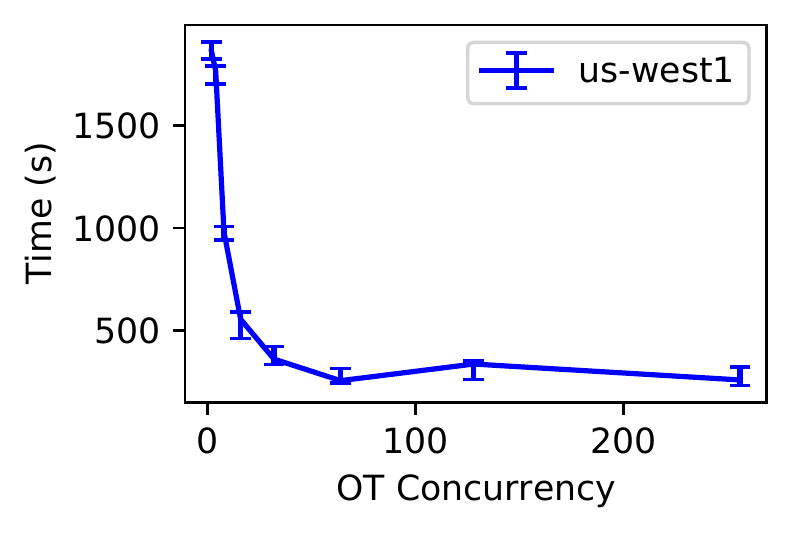}
        \caption{Time to run \merge{} vs. number of concurrent OTs}
        \label{fig:wan_ots}
    \end{subfigure}
    \begin{subfigure}[t]{0.495\linewidth}
        \centering
        \includegraphics[width=\linewidth]{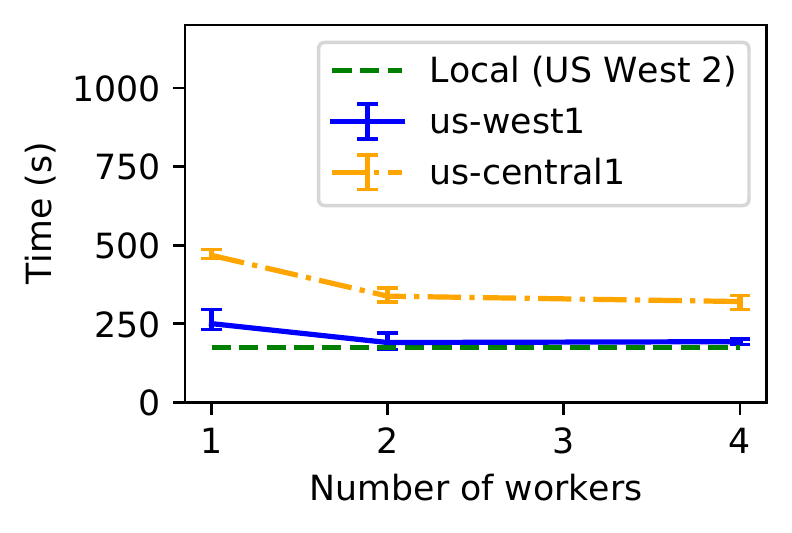}
        \caption{Time to run \merge{} vs. number of workers}
        \label{fig:wan_conn}
    \end{subfigure}
    \caption{Wide-area garbled circuit performance in \sys{}}
    \label{fig:wan}
\end{figure}

\subsection{SMPC in Wide-Area Networks}\label{s:wan}
SC does not always require significant data transfer over the wide area.
In HE, computation is done by a single logical party.
Even in SMPC, there may be ways for multiple parties to co-locate for an SMPC computation while remaining physically and logically distinct.
But in some cases, it is desirable to run SMPC over a wide-area network.
We explore this below.

We measure performance of garbled circuits with the two parties hosted on different cloud providers.
The garbler was always on Azure in West US 2 (Washington). 
The evaluator was on Google Cloud (\texttt{n2-highcpu-2}~\cite{gcloudmachine}).
We compare two setups: one where the evaluator was in \texttt{us-west1} (Oregon) and one where it was in \texttt{us-central1} (Iowa).

Initially, higher latencies and limited single-flow bandwidth limited performance.
For example, the round-trip time in the Oregon setup was $\approx$11 ms, which made OTs a bottleneck.

First, we tuned the local TCP stack, increasing the maximum window size to 32 MiB.
Then, we increased the number of OT rounds performed concurrently, pipelining multiple OT rounds over a single connection, which significantly improved performance (\figref{fig:wan_ots}).
Additionally, we explore parallelizing the computation, assigning multiple workers to the same machine, so that multiple TCP flows are used.
The results are in \figref{fig:wan_conn}.
The dashed line at the bottom is the time to run the experiment with both the garbler and evaluator on Azure (taken from \figref{fig:workloads_single}).
For the Oregon setup, we can come close to the Local performance using two flows.
The Iowa setup is more challenging because less bandwidth is available per flow.
Using multiple parallel flows helps, but the performance improvement in the Iowa setup is limited by variation in wide-area flow performance, which induces stragglers.

In both cases, the performance overhead of operating in the wide area is less than the performance overhead of swapping (\figref{fig:workloads_single}), indicating that \sys{}'s techniques confer substantial benefit even in wide-area settings.

\subsection{Applications}\label{s:applications}
For these experiments, we did not use \texttt{cgroup}s to limit RAM.
The OS and MAGE setups ran using all of the available RAM.

\subsubsection{Detecting Password Reuse}
When users reuse a password across multiple websites, they become prone to ``credential stuffing'' attacks, in which an attacker uses a user's password leaked by one site to compromise that user's account on other sites.
To address this problem, sites may wish to identify which of their users reuse their passwords on other sites~\cite{wang2019password}.
Senate~\cite[Query 2 in \S{}2]{poddar2021senate} proposes a protocol for this.
First, the sites arrange to assign user IDs and hash passwords such that they will match \emph{across} sites.
Then, they use SMPC to detect which user IDs are shared between the sites and have the same password hash.
Note that user IDs and password hashes cannot be shared directly, since they are sensitive (the hashes can be reversed).

We write a two-party version of the password reuse program in \sys{}'s DSL for garbled circuits, based on Senate's password reuse program.
Senate uses a different SMPC protocol, so its results are not directly comparable to ours.

\begin{figure}[t]
    \centering
    \includegraphics[width=\linewidth]{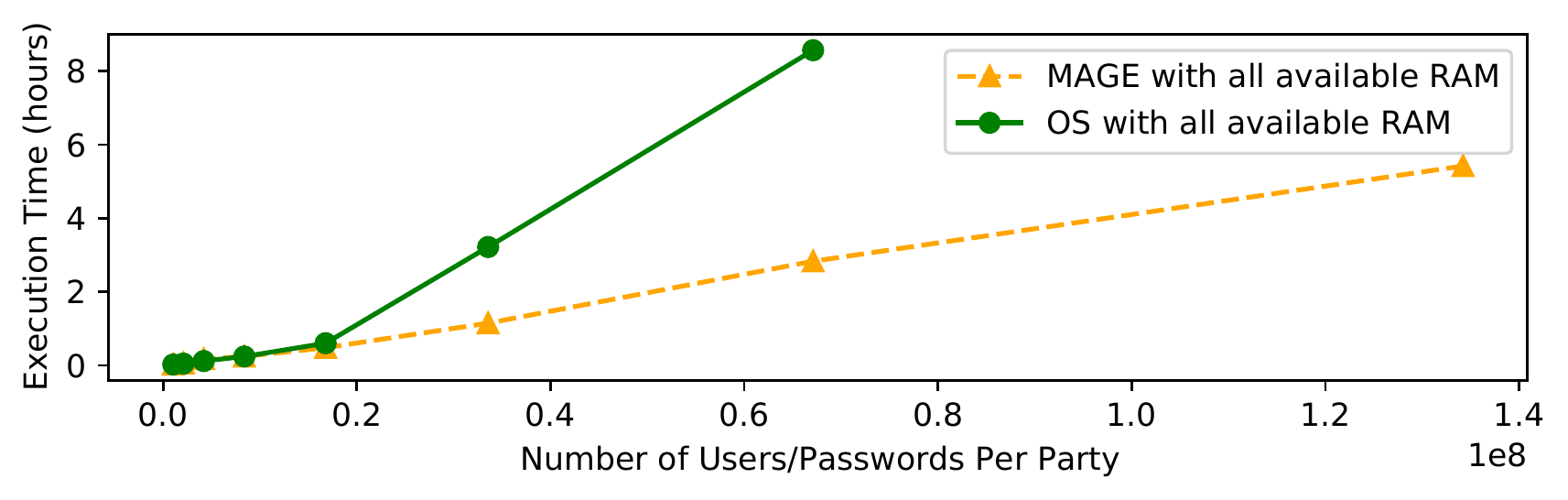}
    \caption{Scaling password reuse detection with \sys{}}
    \label{fig:password_reuse}
\end{figure}

We use \sys{} to scale the password reuse program to $2^{27}$ users per party, which requires 1.125 TiB on each party.
A single \texttt{D16d\_v4} instance does not have enough swap space.
Thus, we use four \texttt{D16d\_v4} instances on Azure for the garbler party, and four \texttt{n2-highmem-4} instances on Google Cloud~\cite{gcloudmachine} for the evaluator party.
As explored in \secref{s:wan}, we use two workers per instance (total of eight workers per party) to efficiently use wide-area network bandwidth.
The results are shown in \figref{fig:password_reuse}.
For a given time budget, \sys{} increases the number of user-password records by $\approx 3\times$.
This improvement may have been larger had we been able to obtain \texttt{Ddv4}-series instances with a greater swap-space-to-RAM ratio.

\begin{figure}[t]
    \centering
    \includegraphics[width=\linewidth]{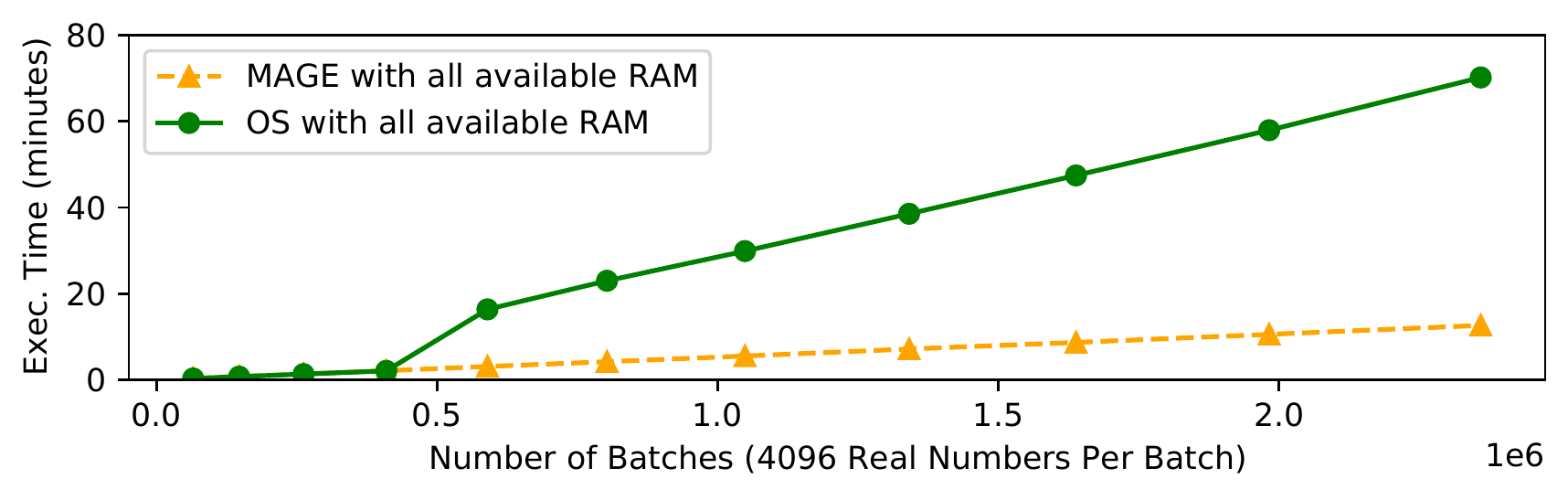}
    \caption{Scaling computational PIR with \sys{}}
    \label{fig:computational_pir}
\end{figure}

\subsubsection{Private Information Retrieval}
Private Information Retrieval (PIR) is a family of protocols that allow a user to retrieve a data item at a particular index from a database without the database learning which item was accessed.
PIR can be used to support private queries on public data~\cite{wang2017splinter}.
We evaluate \sys{} by using CKKS to instantiate the classic Kushilevitz-Ostrovsky single-server computational PIR scheme~\cite[\S{}3]{kushilevitz1997replication}.
PIR's access pattern is particularly simple---a linear scan over the database---so ad-hoc approaches to prefetching, or multi-threading to improve swap performance, may be quite effective.
Our focus is on what \sys{} optimizes \emph{automatically}, so we do not include such ad-hoc optimizations in the OS baseline.
We use a single worker (thread) to compute the PIR.
The database consisted of plaintext data pre-encoded into batches to use with CKKS.
We wrote a DSL program that populates the database (with hardcoded elements) and then performs a PIR query on it; the reported measurements are the time to perform the PIR query, not including the time to populate the database.
The results are in \figref{fig:computational_pir}.
For a given time budget, \sys{} allows for $\approx 5\times$ as many database elements to be processed.

\section{Related Work}\label{sec:related}
Much existing work has looked at high-performance algorithms for SMPC~\cite{wang2017agmpc, damgard2011multiparty, damgaard2013practical, keller2016mascot, keller2018overdrive} and HE~\cite{chillotti2016faster, gentry2012fully}.
These works focus on the cryptography, not how to manage a computer's resources to perform large computations efficiently.

A complementary line of work explores tailoring SMPC computations to a specific application~\cite{juvekar2018gazelle, zheng2019helen, riazi2019xonn, chen2020maliciously}.
The goal of \sys{} is to perform the same computation more efficiently, so its techniques generalize across different applications.
For an application, one may first simplify the computation using application-specific observations, and then execute the resulting computation as efficiently as possible.

Research works including Fairplay~\cite{malkhi2004fairplay}, HEKM~\cite{huang2011faster}, KSS~\cite{kreuter2012billion},
MLB~\cite{mood2012memory}, PCF~\cite{kreuter2013pcf}, and TinyGarble~\cite{songhori2015tinygarble} are frameworks for garbled circuit execution.
We described many of them in \secref{s:memmgmt}.
One work~\cite{buescher2015parallelization} explores parallelizing execution of a garbled circuit, using programming language tools to automatically extract parallelism.
None of them explore how to efficiently swap memory to storage, as \sys{} does.

There already exist many DSLs and compilers for SMPC~\cite{holzer2012cbmc, emp-toolkit, liu2015oblivm, zahur2015oblivc, mood2016frigate, hastings2019compilers, zheng2021cerebro} and HE~\cite{carpov2015armadillo, dathathri2020eva, viand2021compilers}.
These tools often aim to make SC more accessible to non-expert developers, by automatically optimizing the SC program.
\sys{} addresses the complementary problem of executing the resulting SC circuit more efficiently.
To use an existing tool with \sys{} (as in \figref{fig:layering}), one could modify it to output its optimized circuits in one of \sys{}'s DSLs, and then run \sys{}'s planner on that DSL code.
Alternatively, one could modify the tool to output a bytecode directly usable by \sys{}'s planner (e.g., the ``Virtual Bytecode'' in \figref{fig:planner}).

AIFM~\cite{ruan2020aifm} uses similar C++ language features as MAGE's DSLs.
AIFM uses them at runtime for fine-grained memory management.
In contrast, \sys{} (1) executes DSL programs only to extract the memory access pattern during the planning phase and (2) manages memory at the granularity of pages.

There is an extensive literature concerning memory management in traditional operating systems~\cite{belady1966study, denning1968thrashing, belady1969anomaly, denning1970virtualmemory, denning1980workingsets}.
A related line of work looks at how operating systems can give memory-intensive applications, such as scientific simulations, more control over paging~\cite{harty1992external}.
While these works focus primarily on paging in the classic sense, our work explores memory programming.
Additionally, our work, unlike scientific simulations, is capable of \emph{general} computations within SC.
Scheduling page movement according to real-time constraints imposed by computation also draws from the real-time scheduling literature~\cite{liu1973scheduling}.
These techniques do not manage memory directly and are complementary to ours.

Some systems in other domains, like neural network training, formulate memory management problems as an integer linear program and use an exponential-time solver~\cite{jain2020checkmate}.
This approach exploits the high-level structure of the application to coarsen the dataflow graph.
For \sys{}, the dataflow graph is much larger because \emph{general} SC computations do not conform to any particular high-level structure.
By operating on a program representation of the circuit (\secref{s:ir}), \sys{} does coarsen the graph, but it nevertheless remains enormous.
Thus, we use our staged approach (\secref{sec:design}) to find a good approximation.

Some systems use observations of past memory accesses or past working sets (e.g., from prior invocations of a program) to perform targeted prefetching~\cite{zhang2011fast, he2013io, hashemi2018learning, maruf2020leap, ustiugov2021benchmarking} and approximate Belady's algorithm (MIN)~\cite{song2020learning}.
SC's obliviousness and our memory programming approach allow \sys{} to compute the memory access pattern without first running the program, and then apply these techniques using the access pattern itself.

The recent DEMAND-MIN~\cite{jain2018rethinking} algorithm combines MIN with prefetching.
DEMAND-MIN tells which item to evict given an access pattern sequence and prefetch sequence fixed in advance.
It is not directly applicable to \sys{} because \sys{}'s prefetch sequence is not fixed in advance.

At a technical level, \sys{}'s planning is similar to register allocation in compiler
theory~\cite{chaitin1981coloring, cooper1998liverangesplitting, traub1998linearscan, wimmer2005linearscansplitting}---variables, registers, and memory in register allocation correspond to wire values, slots in memory, and storage swap space in the context of \sys{}.
The key difference is that register allocators must deal with conditional branches whose outcomes cannot be predicted at compile time.
From the perspective of register allocation, the entire circuit that \sys{} operates on would be viewed as a single basic block.
We discussed a result from register allocation theory for a single basic block in \secref{s:replacement}.
Another result is that, for a \emph{fixed} number of registers, there is a linear-time algorithm that can reorder instructions within a structured program to optimize its register allocation~\cite[\S{}3.2]{bodlaender1998reordering} (though the time is exponential in the number of registers).

\section{Conclusion}\label{sec:conclusion}
This paper explores how to efficiently execute SC computations that do not fit in memory.
Our key observation is that SC is inherently oblivious.
This enables memory programming, in which one computes the access pattern of an SC program in advance and uses it to produce a memory management plan.
By using memory programming to preplan data transfers between memory and storage, \sys{} runs SC up to an order of magnitude faster than the OS virtual memory system and can execute some SC programs at nearly in-memory speeds.

Some non-SC programs, like plaintext neural network inference and programs designed for hardware enclaves like Intel SGX, are also oblivious.
Applying memory programming to such workloads is an interesting direction for future work.

\section*{Acknowledgments}

We thank the anonymous reviewers and our shepherd, Nadav Amit, for their helpful feedback.
We would also like to thank Katerina Sotiraki and other students/postdocs from the RISELab Security Group for their feedback on early drafts.

This work is supported by NSF CISE Expeditions Award CCF-1730628, NSF CAREER 1943347, and gifts from the Sloan Foundation, Bakar Fellows Program, Alibaba, Amazon Web Services, Ant Group, Ericsson, Facebook, Futurewei, Google, Intel, Microsoft, Nvidia, Scotiabank, Splunk, and VMware.
This research is also supported in part by the National Science Foundation Graduate Research Fellowship Program under Grant No. DGE-1752814.
Any opinions, findings, and conclusions or recommendations expressed in this material are those of the authors and do not necessarily reflect the views of the National Science Foundation.

%
%

\bibliographystyle{plain}
\bibliography{mage}

\appendix
\section{Artifact Appendix}

\subsection*{Abstract}


Our artifact consists of a \sys{} prototype and scripts to use it to run our experiments from \secref{sec:evaluation}.
The \sys{} prototype can execute SC efficiently even when the computation does not fit in memory.
It does so by using memory programming to provide a very efficient virtual memory abstraction.
Our prototype supports distributing an SC computation across workers that communicate over the network, allowing for parallel and distributed SC execution.
The \sys{} prototype presently supports two SC protocols: garbled circuits and CKKS.
It follows the layered architecture described in \secref{s:architecture}.

\subsection*{Scope}


Our artifact can be used to validate our central claim that, using memory programming, \sys{} can execute SC computations that do not fit in memory at nearly in-memory speeds.
Specifically, our artifact can be used to validate the performance claims made in the figures and table in \secref{sec:evaluation}.
Our submitted artifact package allowed the artifact evaluation committee to reproduce those results present in our submitted paper; we have since added support for reproducing the measurements we have added since the original submission.

Our artifact can also be used to run SC computations unrelated to our evaluation of \sys{} in \secref{sec:evaluation}.
The user can describe a custom SC computation using a DSL internal to C++, and then use our \sys{} prototype to generate a memory program for it and execute it efficiently.

\subsection*{Contents}


Our artifact comprises (1) a prototype of \sys{} and (2) scripts to run experiments from \secref{sec:evaluation}.

\medskip
\parhead{Prototype} Our \sys{} prototype includes:
\begin{itemize}[leftmargin=*,noitemsep,topsep=0ex]
    \item The planner and interpreter for the \sys{} system.
    \item A utility program to read the bytecode format used by our implementation and print a memory program in human-readable form.
    \item Implementations of the workloads used in our evaluation (\secref{s:workloads}) in \sys{}'s DSLs.
    \item Utility programs to prepare inputs for these workloads.
    \item A wiki page that walks the user through using our \sys{} prototype to perform a computation.
\end{itemize}

\medskip
\parhead{Scripts} Our scripts to run our experiments include:
\begin{itemize}[leftmargin=*,noitemsep,topsep=0ex]
    \item A program, \texttt{magebench.py}, that can spawn cloud instances on Microsoft Azure and Google Cloud and run experiments on the resulting cloud setup. The command line parameters passed to this program can be used to specify the cloud setup and experiments to run; the user can change these command line parameters to change aspects of the setup (e.g., number of workers, memory size, problem size, etc.).
    \item A README file that describes how to use \texttt{magebench.py} to run our experiments from \secref{sec:evaluation} and obtain log files containing the results.
    \item An IPython notebook to produce graphs from the log files output by \texttt{magebench.py}.
    \item Utility scripts to help automate invoking \texttt{magebench.py} to run experiments from \secref{sec:evaluation}.
\end{itemize}

\subsection*{Hosting}


Our artifact is available on GitHub.
Our \sys{} prototype is available at \url{https://github.com/ucbrise/mage} and our scripts to run our experiments are available at \url{https://github.com/ucbrise/mage-scripts}.
The version we provided to the artifact evaluation committee is marked in both repositories using the \texttt{osdi21ae} tag.
However, we encourage users to use the latest versions of each repository (on the \texttt{main} branch), as they include the newest features and bug fixes, including scripts for additional experiments in \secref{sec:evaluation}.

\subsection*{Requirements}


We developed and tested our artifact on Intel x86-64 systems running Ubuntu 20.04.
We used \texttt{clang++} 10.0.0 to compile our \sys{} prototype.
The \texttt{magebench.py} script spawns cloud instances with an environment appropriate for building and running our \sys{} prototype.
Spawning those cloud instances requires a subscription to Microsoft Azure and Google Cloud.
The particular software dependencies for our artifact are specified in the README files of our two GitHub repositories.

\subsection*{Workflow}


To use our \sys{} prototype, the user first writes a configuration file in YAML describing the execution setup (e.g., network information and swap file for each worker, number of concurrent OTs for garbled circuits, etc.).
For SMPC, information needed only by other parties (e.g., the swap file for other parties' workers) can be omitted from the configuration file.
Next, the user writes a program in a DSL internal to C++ specifying the computation to run.
Then, the user runs \sys{}'s planner, which accepts the DSL program and configuration file as input, for each worker the user will run, and outputs a file containing a memory program for each worker.
The user prepares a file for each worker describing that worker's input for the computation.
Finally, the user runs \sys{}'s interpreter for each worker, which accepts files containing the memory program, configuration, and input data and writes a file containing the program's output.
Further details are given in the README file and wiki pages of the \texttt{mage} repository on GitHub.

To use our script to run experiments, the user invokes \texttt{magebench.py} to spawn cloud virtual machines.
The user can then invoke \texttt{magebench.py} to run \sys{} on those cloud virtual machines, copy the resulting log files to the machine where \texttt{magebench.py} is run, and finally, deallocate the cloud virtual machines.
Further details are given in the README file of the \texttt{mage-scripts} repository on GitHub.

%

\end{document}